\documentclass[12pt]{article}
\usepackage{latexsym}
\font\tenmsbm=msbm10 scaled 1200
\font\sevenmsbm=msbm9
\newfam\msbmfam
\textfont\msbmfam=\tenmsbm
\scriptfont\msbmfam=\sevenmsbm

\def\msbm{\fam\msbmfam\tenmsbm}

\makeatletter
\@addtoreset{equation}{section}
\makeatother
\renewcommand{\theequation}{\thesection.\arabic{equation}}
\newcounter{parentequation}
\newenvironment{subequations}{%
  \refstepcounter{equation}%
  \begingroup 
  \let\protect\noexpand
  \edef\@tempa{\def\noexpand\theparentequation{\theequation}}%
  \expandafter
  \endgroup\@tempa
  \setcounter{parentequation}{\value{equation}}%
  \setcounter{equation}{0}%
  \def\theequation{\theparentequation\alph{equation}}%
  \ignorespaces
}{%
  \setcounter{equation}{\value{parentequation}}%
}
\newcommand{\eqn}[1]{(\ref{#1})}
\newsavebox{\uuunit}
\sbox{\uuunit}
    {\setlength{\unitlength}{0.825em}
     \begin{picture}(0.6,0.7)
        \thinlines
        \put(0,0){\line(1,0){0.5}}
        \put(0.15,0){\line(0,1){0.7}}
        \put(0.35,0){\line(0,1){0.8}}
       \multiput(0.3,0.8)(-0.04,-0.02){10}{\rule{0.5pt}{0.5pt}}
     \end {picture}}
\newcommand {\unity}{\mathord{\!\usebox{\uuunit}}}
\newsavebox{\bobox}
\sbox{\bobox}
    {\setlength{\unitlength}{0.825em}
     \begin{picture}(0.6,0.7)
        \thinlines
        \put(0,0){\line(1,0){0.7}}
        \put(0,0){\line(0,1){0.7}}
        \put(0.7,0){\line(0,1){0.7}}
        \put(0,0.7){\line(1,0){0.7}}
        \multiput(0,0)(0.05,0.05){14}{\rule{0.4pt}{0.4pt}}
        \multiput(0,0.65)(0.05,-0.05){13}{\rule{0.4pt}{0.4pt}}
     \end {picture}}
\newcommand {\boxtimes}{\mathord{\!\usebox{\bobox}}\,}
\def\IC{\relax\,\hbox{$\inbar\kern-.3em{\rm C}$}}
\def\bfzero{\relax\,\hbox{$\inbar\kern-.3em{\rm 0}$}}

\def\bfone{\relax{\rm 1\kern-.35em 1}}
 \def\cB{{\cal B}}
 \def\cD{{\cal D}}

 \def\cG{{\cal G}}

\def\cN{{\cal N}} \def\cO{{\cal O}}

\def\beq{\begin{equation}}
\def\eeq{\end{equation}}
\def\bea{\begin{eqnarray}}
\def\eea{\end{eqnarray}}
\def\bet{\begin{tabular}}
\def\eet{\end{tabular}}
\def\bes{\begin{subequations}\bea}
\def\ees{\eea\end{subequations}}
\newcommand{\dfrac}{\displaystyle \frac}

\def\a{\alpha}
\def\b{\beta}
\def\th{\vartheta}
\def\bth{\bar{\vartheta}}
\def\da{\dot{\alpha}}
\def\db{\dot{\beta}}

\def\s{\sigma}
\def\e{\epsilon}

\def\l{\lambda}
\def\hT{\hat{T}}
\def\IC{\hbox{\msbm C}}
\begin{document}

\begin{titlepage}

\begin{flushright}
CERN--TH/99--156 \\
DFTT 99/29 \\
LPTENS 99/19 \\
hep-th/9905226\\
\end{flushright}

\vspace{2truecm}

\begin{center}

{ \Large \bf Spectrum of Type IIB Supergravity on $AdS_5 \times T^{11}$:
Predictions on $\cN =1$ SCFT's }

\vspace{1cm}

{ Anna Ceresole$^{\star}$}\footnote{ceresole@athena.polito.it},
{Gianguido Dall'Agata$^{\dag}$}\footnote{dallagat@to.infn.it},
{Riccardo D'Auria$^{\star}$}\footnote{dauria@polito.it}
and
{Sergio Ferrara$^{\S}$}\footnote{sergio.ferrara@cern.ch}

\vspace{1cm}

{ $\star$ \it Dipartimento di Fisica, Politecnico di Torino  \\
C.so Duca degli Abruzzi, 24, I-10129 Torino, and\\
Istituto Nazionale di Fisica Nucleare, Sezione di Torino. \\
}

\medskip

{$\dag$ \it Dipartimento di Fisica Teorica, Universit\`a  di Torino and \\
Istituto Nazionale di Fisica Nucleare, Sezione di Torino, \\
 via P. Giuria 1, I-10125 Torino.}

\medskip

{$\S$ \it TH Division, CERN, 1211 Geneva 23, Switzerland and \\
Ecole Normale Superieure, Laboratoire de Physique Th\'eorique,\\
    24 rue Lhomond, F-75231 Paris CEDEX 05, France. \\
}

\vspace{1cm}

\begin{abstract}
We derive the full Kaluza--Klein spectrum of type IIB supergravity
compactified on
$AdS_5 \times T^{11}$ with $T^{11} = \dfrac{SU(2) \times SU(2)}{U(1)}$.
From the knowledge of the spectrum and general multiplet shortening
conditions, we make a refined test of the $AdS/CFT$ correspondence, by
comparison between various shortenings of $SU(2,2|1)$
supermultiplets on $AdS_5$
and different families of boundary operators with protected
dimensions.
Additional towers of long multiplets with rational dimensions, that are
not protected by supersymmetry, are also predicted from the supergravity
analysis.
\end{abstract}

\end{center}

\vskip 0.5truecm


\end{titlepage}

\newpage

\baselineskip 6 mm

\section{Introduction}

One of the most fascinating properties of the $AdS/CFT$ correspondence
\cite{M,GKP,W} is the
deep relation between supergravity and gauge theory dynamics, at
least in the regime where the supergravity approximation (small space--time
curvature) is a reliable description of a more fundamental theory such as
string or M theory \cite{AdSReview,FZ}.

Although many tests have been performed in the case of maximal supersymmetry,
relating for instance, the dynamics of $N$ coincident D3 branes
(for large $N$) and
type IIB supergravity compactified on $AdS_5\times S^5$ \cite{AdSReview},
much less is known on
the dual theories for a lower number of supersymmetries \cite{KS}, where the
candidate models exhibit a far richer structure since they contain a variety
of matter  multiplets with additional symmetries other than the original
$R$-symmetry dictated by the supersymmetry algebra \cite{Keh}.

A particularly interesting class of models are obtained by assuming that $S^5$
is replaced by a five--dimensional coset manifold $X_5=G/H$ with some Killing
spinors. As shown in \cite{Rom} there is a unique such manifold
$X_5=T^{pq}=\dfrac{SU(2)\times SU(2)}{U(1)}$ with $p=q=1$, where $p$ and $q$
define the embedding of the $H=U(1)$ group into the two $SU(2)$ groups  .
 The supergravity
theory on $AdS_5 \times T^{11}$ is an $\cN=2$ supergravity theory with a matter
gauge group $G=SU(2)\times SU(2)$.
The corresponding four dimensional conformal field theory  must then be
\cite{KW} an $\cN=1$
Yang--Mills theory with a flavour symmetry $G$ such that an accurate test
of the $AdS/CFT$ correspondence could be made using the knowledge of the entire
spectrum of the supergravity side of this theory.

The conformal field theory description of $IIB$ on $AdS_5\times T^{11}$ was
constructed by Klebanov and Witten \cite{KW}
and it was the first example of a conformal theory describing branes at
conifold singularities.
The same theory was later re--obtained by Morrison and Plesser \cite{MP}
by adopting a general method of studying
branes at singularities \cite{conifolds}.
Infact, under certain conditions, a conical singularity in a Calabi--Yau space of
complex dimension $n$ can be described by a cone over an Einstein manifold
$X_{2n-1}$. In the case of $X_5=T^{11}$ such construction gives rise to a
conformal field theory with ``singleton'' \cite{FF} degrees of freedom $A$ and
$B$ each a doublet of the  factor groups $SU(2)\times SU(2)$ and with
conformal anomalous dimension $\Delta_{A,B}=3/4$. Moreover the gauge group $\cG$ is
$SU(N)\times SU(N)$ and the two singleton (chiral) multiplets are
respectively in the $(N,\overline N)$ and $(\overline N,N)$ of $\cG$.

An infinite set of chiral operators of this theory which are the analogue of
the Kaluza--Klein (KK) excitations of the $\cN=4$ Yang--Mills theory with $SU(N)$
gauge group is given by ${\it Tr}(AB)^k$ with $R$--charge $k$ and in the
($\frac{k}{2}$,$\frac{k}{2}$) representation of $SU(2)\times SU(2)$.
The existence of this
infinite family of chiral operators (massive $\cN=2$ hypermultiplets in the
supergravity language ) has been confirmed by Gubser \cite{G} by a study of
the eigenvalues of the scalar Laplacian when performing harmonic analysis of
IIB supergravity on $AdS_5\times T^{11}$.

Moreover the matching of gravitational and R-symmetry anomalies in the two
theories have been also proved in ref. \cite{G}.

This paper analyses the complete spectrum of the KK
states on $AdS_5\times T^{11}$ and infers its multiplet structure
as done in previous investigations for  maximal supersymmetry.
In that case the KK spectrum, analysed in terms of AdS representations
in \cite{GM,GMZ}, was interpreted in terms of $\cN=1$ conformal
superfields in \cite{W} and in terms of the $\cN=4$ one in
\cite{FFZ} and \cite{AF}.
The multiplet shortening conditions \cite{multsh}
can be inferred from the knowledge of all the mass matrices in the KK spectrum
\cite{CFN,M111}.
In the case of the $SU(2,2|1)$ superalgebra, the shortening is proven to
correspond to three types of shortening of the appropriate
representations, as discussed in \cite{FZ2} and \cite{FGPW}: massless $AdS$
multiplets, short $AdS$ multiplets and  semi--long $AdS$ multiplets.
These multiplets, in the conformal field theory language,
correspond to respectively
conserved, chiral and semi--conserved superfields
which have all protected dimensions and which therefore
correspond to very particular shortening conditions in the KK context.

We show a full and detailed correspondence between all the CFT operators and the
KK modes for the conformal operators of preserved scaling dimension.
We also show that there exist other operators related to long
multiplets but having non--renormalised conformal dimension.
Interestingly enough, these operators seem to be the lowest dimensional
ones for a given structure appearing in the supersymmetric Born--Infeld
action of the $D3$--brane on $AdS_5\times T^{11}$ \cite{ts,GKH,DT,marzaf}.

The paper is organised as follows. In section 2 the harmonic analysis of IIB
supergravity on $AdS_5\times T^{11}$ is performed and the complete mass
spectrum of the theory is exhibited. In sec. 3 properties of $\cN=1$
four--dimensional supersymmetric field theories are recalled, in particular the
superfield realisation of different short and long superconformal multiplets
of the $SU(2,2\vert 1)$ superalgebra.
In sec. 4 a comparison of superfields of protected dimensions and states in the
KK spectrum is made using the formulae giving the mass--conformal dimension
relations as predicted by the $AdS/CFT$ correspondence.

\section{Harmonic analysis on $T^{11}$}

In this section we give a summary of the derivation of the full mass spectrum
of Type IIB supergravity compactified on $AdS_5 \times T^{11}$ obtained by
KK harmonic expansion on $T^{11}$. Since our main goal here is the
comparison of the mass spectrum with the composite operators of the CFT at the
boundary of $AdS_5$, we just sketch the general procedure and postpone  a
detailed derivation of our results to a forthcoming pubblication \cite{CDD}.
Partial results were obtained in \cite{G,JR} using different methods.

\subsection{$T^{11}$ geometry}

Let us start with a short discussion of the $T^{11}$ geometry\footnote{
For details about the notations and conventions see the appendix.}.
We consider two copies of $SU(2)$ with generators $T_A$, $\hT_A$,
($A=1\ldots3$):
\beq
\, [T_A,T_B] = \e_{AB}{}^C T_C, \qquad
\, [\hT_A,\hT_B] = \e_{AB}{}^C \hT_C.
\eeq
We decompose the Lie algebra {\msbm G} of $SU(2) \times SU(2)$ with respect
to the diagonal generator
\beq
\label{TH}
T_{H} \equiv T_3 + \hT_3,
\eeq
as
\beq
\hbox{\msbm G}=\hbox{\msbm H}+\hbox{\msbm K},
\eeq
where the sub--algebra {\msbm H} is made of the single generator $T_H$
and the coset algebra {\msbm K} contains
the generators $T_i$ ($i=1,2$), $\hT_{s}$ ($s = 1,2$), and
\beq
\label{T5}
T_5 = T_3 - \hT_3.
\eeq
In terms of this new basis the commutation relations are
\bea
\, [T_i,T_j] = \frac{1}{2} \e_{ij} (T_H + T_5), &&
\, [\hT_{s},\hT_{t}] = \frac{1}{2} \e_{st} (T_H - T_5), \nonumber \\
\label{Alg}
\, [T_5,T_i]=[T_H,T_i] = \e_i{}^j T_j, &&
\, [T_5,\hT_{s}]=[T_H,\hT_{s}] = \e_{s}{}^{t} \hT_{t}, \\
\, [T_i,\hT_{s}] &=& [T_5,T_H] = 0. \nonumber
\eea
We introduce the coset representative $L$ of $\dfrac{SU(2)\times
SU(2)}{U_H(1)}$, $U_H(1)$ being the diagonal subgroup of $G$
generated by $T_{H}$
\beq
L(y^i,y^s,y^{5})= exp \, T_i y^i \cdot  exp \, \hT_{s} y^{s} \cdot  exp \, T_5
y^5,
\eeq
and construct the left invariant form on the coset
\beq
L^{-1}dL = \omega^i T_i + \omega^{s} \hT_{s} + \omega^5 T_5 +
\omega^H T_H,
\eeq
where the one--forms \{$\omega^i,\omega^{s},\omega^5,\omega^H$\}
satisfy the Maurer--Cartan equations
\beq
\label{MCE}
d \omega^{\Lambda} + \frac{1}{2} C^{\Lambda}{}_{\Sigma\Pi} \;
\omega^{\Sigma} \, \omega^{\Pi} = 0, \qquad
\Lambda,\Pi,\Sigma\equiv\{i,s,5,H\}.
\eeq
The one--forms $\omega^{K} \equiv \{\omega^i,\omega^s,\omega^5\}$
are {\msbm K}--valued  and can be identified with the five vielbeins of
$G/H = T^{11}$, while $\omega^H$ is {\msbm H}--valued  and is called the
$H$--connection of the coset manifold.
It is convenient to rescale the $\omega^K$ and define as
vielbeins $V^a\equiv (V^i,V^{s},V^5)$:
\beq
V^i = a \,\omega^i, \qquad
V^{s} = b\, \omega^s,  \qquad
V^5 = c \, \omega^5,
\eeq
where $a,b,c$ are real rescaling factors which will be determined by
requiring that $T^{11}$ is an Einstein space \cite{Cast,libro}.

Once we have the vielbeins, we may construct the Riemann connection
one--form $\cB^{ab} \equiv -\cB^{ba}$ ($a,b = i,s,5$), imposing the
torsion--free condition
\beq
\label{zerT}
d V^a - \cB^{ab} V_b = 0.
\eeq
By comparison with the M.C.E.'s \eqn{MCE}, one finds
\beq
\label{B}
\begin{array}{lcl}
\cB^{ij} = -\e^{ij} \left[ \omega + \left(c-\dfrac{a^2}{4c}\right)V^5\right], &&
\cB^{5i} = \dfrac{a^{2}}{4c} \, \e^{ij} \, V_j,  \\
\cB^{st} = -\e^{st} \left[ \omega - \left(c-\dfrac{b^2}{4c}\right)V^5\right], &&
\cB^{5s} = -\dfrac{b^{2}}{4c} \, \e^{st}\, V_{t}.  \\
\cB^{is} = 0.&&
\end{array}
\eeq
Consequently, the curvature two--form,  defined as
\beq
R^{ab} \equiv d \cB^{ab} - \cB^a{}_c \cB^{cb},
\eeq
turns out to be
\bea
R^{ij} &=& \left(a^2 - \frac{3}{16} \frac{a^4}{c^2}\right) \, V^iV^j +
\frac{a^2b^2}{16c^2} \; \e^{ij} \e^{st} \, V_{s}V_{t}, \nonumber \\
R^{st} &=& \left(b^2 - \frac{3}{16} \frac{b^4}{c^2}\right) V^{s}V^{t} +
\frac{a^2b^2}{16c^2}  \; \e^{st} \e^{ij} \, V_{i}V_{j}, \nonumber \\
R^{is} &=& \frac{a^2 b^2}{16 c^2} \; \e^{ij} \e^{st} V_j V_{t}, \\
R^{i5} &=& \frac{a^4}{16 c^2} \; V^i V^5, \nonumber\\
R^{s5} &=& \frac{a^4}{16 c^2} \; V^{s} V^5. \nonumber
\eea
The Ricci tensors are now easily computed.
We find
\beq
{R^i}_{k} = \left(\frac{1}{2} a^2 - \frac{a^{4}}{16c^2}\right)\delta^i_k,
\qquad
{R^{s}}_{t} = \left(\frac{1}{2} b^2 -
\frac{b^{4}}{16c^2}\right)\delta^{s}_{t},
\qquad
{R^5}_{5} = \frac{a^4}{8c^2}.
\eeq

In order to have an Einstein space with Ricci tensor
\beq
R^{a}{}_{b} = 2 \, e^2 \, \delta^a_b,
\eeq
we must have
\beq
\label{abc}
a^2 =b^2 = 6 e^2, \quad  \hbox{ and } \quad
c^2 = \frac{9}{4} e^2.
\eeq

\subsection{Harmonic calculus}

An essential tool for the computation of the Laplace--Beltrami invariant
operators on $T^{11}$ is the evaluation of the covariant derivative
$\cD \equiv (\cD_i,\cD_{s},\cD_5)$.
Starting from the definition
\beq
\cD = d + \cB^{ab}T_{ab} \equiv d+\cB,
\eeq
where $T_{ab}$ are the $SO(5)$ generators written as matrices:
$ (T_{ab})^{cd} = - \delta^{cd}_{ab},$
setting $\cB = \omega^H + M$, one can write
\beq
\label{deco}
\cD = \cD^H + M,
\eeq
where the H--covariant derivative is defined by
\beq
\label{DH}
\cD^H =  d + \omega^H
\eeq
and the matrix of one--forms $M$ can be computed from \eqn{B}
\bea
M^{ij} = -\left(c-\frac{a^2}{4c}\right) V^5 \e^{ij}, &&
M^{5i} = \frac{a^{2}}{4c} \e^{ij} V_j, \nonumber \\
\label{M}
M^{st} = \left(c-\frac{a^2}{4c}\right) V^5 \e^{st}, &&
M^{5s} = -\frac{a^{2}}{4c} \e^{st} V_{t}, \\
M^{is} &=& 0. \nonumber
\eea

The usefulness of the decomposition \eqn{deco}, \eqn{DH}, \eqn{M}
lies in the fact that the
action of $\cD^H$ on the basic harmonic represented by the $T^{11}$
coset representative $L^{-1}$ can be computed algebraically.
Indeed one has quite generally \cite{libro,SS}
\beq
\cD^H = - r(a) T_a V^a \equiv -a (T_i V^i + \hT_{s}V^{s})- c T_5 V^5,
\eeq
where $r(i) = r(s) = a$, $r(5) = c$ are the rescalings
and $T_a$ are the coset generators of $T^{11}$.

In summary, the covariant derivative on the basic harmonic $L^{-1}$
can be written as follows
\beq
\label{DL-1}
\cD L^{-1} = (-r(a) T_a V^a + M^{ab}T_{ab}) L^{-1},
\eeq
or, in components, using \eqn{M},
\bea
\cD_i L^{-1} &=& \left(-a T_i - \frac{a^2}{2c} \e_i{}^j T_{5j}\right) L^{-1}, \nonumber\\
\label{DaL-1}
\cD_{s} L^{-1} &=& \left(-a T_s + \frac{a^2}{2c} \e_{s}{}^{t}
T_{5t}\right) L^{-1},\\
\cD_5 L^{-1} &=& \left(-c T_5 - 2\left(c-\frac{a^2}{4c}\right)(T_{12} - T_{34})
 \right) L^{-1}. \nonumber
\eea

In a KK compactification, after the linearisation of the
equations of motion of the field fluctuations,
one is left with a differential equation
on the ten--dimensional fields $\phi_{[\l_1,\l_2]}^{[\Lambda]}
(x,y)$
\beq
\label{LB}
(\Box^{[\Lambda]}_x + \boxtimes^{[\l_1,\l_2]}_y)
\phi^{[\Lambda]}_{[\l_1,\l_2]}(x,y) = 0.
\eeq
Here the field $\phi_{[\l_1,\l_2]}^{[\Lambda]} (x,y)$ transforms irreducibly in
the representations $[\Lambda] \equiv [E_0,s_1,s_2]$ of $SU(2,2)
\approx O(4,2)$ and $[\l_1,\l_2]$ of $SO(5)$ and it depends on the
coordinates $x$ of $AdS_5$ and $y$ of $T^{11}$.
$\Box_x$ is the kinetic operator for a field of quantum number $[\Lambda]$
in five--dimensional $AdS$ space and $\boxtimes_y$ is the
kinetic operator for a field of spin $[\l_1,\l_2]$  in
the internal space $T^{11}$.
(In the following we omit the index $[\Lambda]$ on the fields).

Expanding $\phi_{[\l_1,\l_2]} (x,y)$ in the harmonics of
$T^{11}$ transforming irreducibly under the isometry group of
$T^{11}$, one is reduced to the problem of computing the action of
$\boxtimes_y$ on the harmonics, whose eigenvalues define the $AdS$ mass.

$\boxtimes_y$ is a Laplace--Beltrami operator on $T^{11}$
and it is constructed, for every representation
$[\l_1,\l_2]$, in terms of the covariant derivative on $G/H$.
Since the covariant derivative acts
algebraically on the basic vector or spinor
harmonic $L^{-1}$ (in terms of which {\it any}
harmonic can be constructed), the
problem of the mass spectrum computation is reduced, via
\eqn{DL-1}--\eqn{DaL-1} to a purely algebraic problem.

The explicit evaluation of the linearised equation \eqn{LB} for the
five--dimensional case has been given
in \cite{KRV} and we will adopt the same notations therein to
denote the five--dimensional space--time fields appearing in the
harmonic expansion.
Note that \eqn{LB} has been evaluated in \cite{KRV} around the background
solution presented in \cite{Rom}:
\bea
F_{abcde}=  e \e_{abcde}, &&{R^a}_b = 2 e^2 \delta^a_{b}, \nonumber \\
\label{Romm}
F_{mnpqr}=  -e \e_{mnpqr}, &&{R^m}_n = -2 e^2 \delta^m_{n}, \\
B = A_{MN} = 0, &\qquad&\psi_{M}=\chi=0, \nonumber
\eea
where the field $F_{abcde}$ and $F_{mnpqr}$ is the projection
on $T^{11}$ and $AdS_5$  of the
ten--dimensional five--form $F$ defined as $F=dA_4$, $A_4$ being the
real self--dual four--form of type IIB supergravity.
The other fields of type $IIB$ supergravity are:
the metric $G_{MN}(x,y)$ with internal and space--time components
$g_{\a\beta}(y)$, $g_{\mu\nu}(x)$ whose Ricci tensors in this background
are given in \eqn{Romm} and the complex 0--form and 2--form $B$ and
$A_{MN}$ (the fermionic fields $\psi_M$ and $\chi$ are obviously zero
in the background \eqn{Romm}).

\subsection{Harmonic expansion}

The harmonics on the coset space $T^{11}$
are labelled by two kinds of indices, the first labelling the
particular representation of the isometry group $SU(2) \times SU(2)
\times U_R(1)$ and the other referring to the representation of the
subgroup $H \equiv U_H(1)$.
The  harmonic is thus denoted by $Y_{(q)}^{(j, l, r)} (y)$
where $j$, $l$ are the spin quantum numbers of the two $SU(2)$ in a
given representation, $q$ is the $U_H(1)$ charge and $r$ denotes
the $U_R(1)$ quantum number
associated to the generator $T_5$ orthogonal to $T_H$.
We can identify $r$ as the $R$--symmetry quantum number \cite{G,JR}.

Now we observe that $U_H(1)$ is necessarily a subgroup of $SO(5)$,
the tangent group of $T^{11}$.
The embedding formula of $U_H(1)$ in a given representation of
$SO(5)$ labelled by indices $\Lambda$, $\Sigma$, is given by \cite{libro,SS}
\beq
(T_H)^{\Lambda}{}_{\Sigma} = C_H{}^{ab} (T_{ab})^{\Lambda}{}_{\Sigma},
\eeq
where the structure constants $C_H{}^{ab}$ are derived from the algebra
\eqn{Alg} and $T_{ab}$ are the $SO(5)$ generators.

In the vector representation of $SO(5)$ we find
\beq
\label{THv}
(T_H)_{ab} = C_{Hab} = \left(
\bet{ccc}
$\e_{ij}$ & &\\
 & $\e_{st}$&\\
 & &0
 \eet
\right),
\eeq
while for the spinor representation we get
\beq
\label{THs}
(T_H) = C_H{}^{ab}(T_{ab})=-\frac{1}{4}
C_H{}^{ab}(\gamma_{ab}) = -\frac{1}{2}(\gamma_{12} +
\gamma_{34}) = i \left(
\bet{cccc} 0 &&& \\ &0&& \\ && 1 & \\ &&&-1 \eet
\right),
\eeq
where $\gamma$ are the $SO(5)$ gamma matrices.

The above results imply that an $SO(5)$ field
$\phi_{[\l_1,\l_2]}(x,y)$ can be splitted into the direct sum of
$U_H(1)$ one--dimensional fragments labelled by the $U_H(1)$ charge $q$.
From \eqn{THv} and \eqn{THs} it follows that the five--dimensional
and four--dimensional $SO(5)$ representations break  under
$U_H(1)$ as
\beq
\label{decco}
\begin{array}{rcll}
\hbox{\bf 5} &\to& 1 \oplus -1 \oplus 1 \oplus -1 \oplus 0 &
[\l_1,\l_2] = [1,0],\\
\hbox{\bf 4} &\to& 1 \oplus -1 \oplus 0 \oplus 0 \qquad  &
[\l_1,\l_2] = [1/2,1/2].
\end{array}
\eeq

From \eqn{decco} we easily find the analogous breaking law for
antisymmetric tensors ($[\l_1,\l_2] = [1,1]$), symmetric
traceless tensors ($[\l_1,\l_2] = [2,0]$) and spin tensors
($[\l_1,\l_2] = [3/2,1/2]$) by taking suitable combinations:
\beq
\begin{array}{rcll}
\hbox{\bf 10} &\to& \pm 1 \oplus \pm 1 \oplus \pm 2 \oplus 0 \oplus 0
\oplus 0 \oplus 0 &
[\l_1,\l_2] = [1,1], \\
\hbox{\bf 16} &\to& \pm 2 \oplus \pm 2 \oplus \pm 1 \oplus \pm 1
\oplus \pm 1 \oplus \pm 1 \oplus 0 \oplus 0 \oplus 0 \oplus 0 &
[\l_1,\l_2] = \left[\frac{3}{2},\frac{1}{2}\right], \\
\hbox{\bf 14} &\to& \pm 2 \oplus \pm 2 \oplus \pm 2 \oplus \pm 1 \oplus \pm 1
 \oplus 0 \oplus 0 \oplus 0 \oplus 0 & [\l_1,\l_2] = [2,0].
\end{array}
\eeq

Actually it is often more convenient to write down the harmonic expansion
in terms of the $SO(5)$ harmonics $Y^{(j,l)}_{[\l_1,\l_2]}$
whose fragments are the $Y^{(j,l,r)}_{(q)}$ introduced before.

The generic field
$\phi_{[\l_1,\l_2]}(x,y)$  can be expanded in these harmonics as follows
\beq
\label{hexp}
\phi_{ab\ldots}(x,y) = \sum_{(\nu)} \sum_{(m)} \phi_{(\nu)(m)}(x)
Y^{(\nu)(m)}_{ab\ldots} (y),
\eeq
where $a,b,\ldots$ are $SO(5)$ tensor (or spinor) indices of the
representation $[\l_1,\l_2]$, $(\nu)$ is a shorthand notation for $(j,l,r)$
and $m$ labels the representation space of ($j,l,r$).
In our case $m$ coincides with the labelling of the $U_H(1)$ fragments.
It is well known \cite{libro,SS} that the irrepses of $SU(2)  \times SU(2)$
appearing in the expansion \eqn{hexp} are only those which contain, when
reduced with respect to $U_H(1)$, a charge $q$ also appearing in the
decomposition of $[\l_1,\l_2]$ under  $U_H(1)$.

It is easy to see which are the constraints on $j,l,r$ selecting
the allowed representations $(\nu)$ appearing in \eqn{hexp}.
We write a generic representation of $SU(2) \times SU(2)$ in the Young
tableaux formalism:
\beq
\label{Tab1}
(j,l) \equiv \underbrace{\bet{|c|c|c|}\hline &$\ldots$ &
\\\hline\eet}_{2j}
\otimes \underbrace{\bet{|c|c|c|}\hline &$\ldots$ &
\\\hline\eet
}_{2l}.
\eeq
A particular component of \eqn{Tab1} can be written as
\beq
\label{Tab2}
 \underbrace{\bet{|c|c|c}\hline 1 &$\ldots$ & 1 \\\hline\eet}_{m_1}
\underbrace{\bet{|c|c|c|}\hline 2 &$\ldots$ & 2 \\\hline\eet}_{m_2}
\otimes
 \underbrace{\bet{|c|c|c}\hline 1 &$\ldots$ & 1 \\\hline\eet}_{n_1}
\underbrace{\bet{|c|c|c|}\hline 2 &$\ldots$ & 2 \\\hline\eet}_{n_2}
\eeq
and we have
\beq
\left\{ \bet{rcl} $2j$ &=& $m_1+m_2$ \\ $2j_3$ &=& $m_2-m_1$\eet \right. ,
\qquad
\left\{ \bet{rcl} $2l$&=& $n_1+n_2$ \\ $2l_3$ &=& $n_2-n_1$\eet \right. .
\eeq
Furthermore (recalling the definitions \eqn{TH}--\eqn{T5}) we get
\beq
\begin{array}{rcl}
T_H Y_{(q)}^{(j,l,r)} &=&{\rm i} \, q \, Y_{(q)}^{(j,l,r)}  \equiv {\rm
i} \, (j_3 +l_3) \, Y_{(q)}^{(j,l,r)}, \\
T_5 Y_{(q)}^{(j,l,r)} &=& {\rm i} \, r \, Y_{(q)}^{(j,l,r)}  \equiv {\rm i}
\, (j_3 - l_3) \, Y_{(q)}^{(j,l,r)}.
\end{array}
\eeq
Hence
\bea
\label{qr}
\begin{array}{rcl}
2j_3 &=& q+r \equiv m_2-m_1, \\
2l_3 &=& q-r \equiv n_2-n_1.
\end{array}
\eea
Now we observe that as long as $m_2-m_1$ and $n_2-n_1$ are even or
odd, the same is true for $m_1+m_2$ and $n_1+n_2$.
Therefore the parity of $2j$ and $2l$ is the same as that of
$2j_3$ and $2l_3$  and since $2j_3+2l_3= 2q$ can be even or
odd, the same is true for $2j+2l$.
It follows that $j$ and $l$ must either be both integers or
both half--integers.
This means that the $q$ value of any $U_H(1)$ fragment of the $SO(5)$ fields
is always contained in any $SO(5)$--harmonic in the irrep $(j,l)$ provided
that $j$ and $l$ are both integers or
half--integers.
Since $q+r$ and $q-r$ are related to the third component of the "angular
momentum" of the two $SU(2)$ factors, one also has the conditions
$|q+r| \leq 2j$ and $|q-r| \leq 2l$.
The two above conditions select the harmonics
appearing in the expansion.

In order to be specific it is now convenient to list all the five--dimensional
space--time fields appearing in the harmonic expansion together with the
corresponding ten--dimensional fields,  with $AdS_5$ indices and/or internal
indices,  following the notations of \cite{KRV} . We  group  them according to
the appropriate $SO(5)$ bosonic ($Y$) or fermionic ($\Xi$) harmonic .
\begin{table}[ht]
\begin{center}
\vskip .3cm
 \begin{tabular}{|c|c|c|c|c|c|c|}\hline
10 D &$h_{\mu\nu}$ & $h^a{}_a$ &   $A_{abcd}$     &   $B$  &   $A_{\mu\nu}$&           \\
5  D &$H_{\mu\nu}$ & $\pi$             &   $b$            &   $B$  &   $a_{\mu\nu}$& $Y$      \\\hline
10 D &$h_{a\mu}$   & $A_{\mu abc}$     &   $A_{\mu a}$    &        &                &           \\
5  D &$B_\mu$      & $\phi_\mu$        &   $a_\mu$        &        &               & $Y_a$     \\\hline
10 D &$A_{\mu\nu ab}$&$A_{ab}$         &                  &        &               &           \\
5  D &$b^\pm_{\mu\nu}$      &$a$                &                  &        &               & $Y_{[ab]}$\\\hline
10 D &$h_{ab}$     &                   &                  &        &               &            \\
5 D  &$\phi$       &                   &                  &        &               & $Y_{(ab)}$\\\hline
\hline
10D  &$\lambda$    &$\psi_{(a)}$           &$\psi_\mu$        &        &               &          \\
5D   &$\lambda$    &$\psi^{(L)}$         &$\psi_\mu$      &        &               &$\Xi$    \\\hline
10D  &$\psi_a$      &                    &                 &        &               &          \\
5 D  &$\psi^{(T)}$  &                    &                 &        &               &$\Xi_a$  \\
\hline
 \end{tabular}
\caption{Fields appearing in the harmonic expansion.}
\end{center}
\label{fiel}
\end{table}

Note that the ten--dimensional fields $h^{\mu}{}_{\mu}(x,y)$,
$A_{\mu\nu\rho\s}(x,y)$, $A_{\mu\nu\rho a}(x,y)$
are not part of the above list
since, as shown in \cite{KRV}, they appear algebraically in the linearised
equations of motion and thus can be eliminated in terms of the other
propagating fields.

To obtain the mass spectrum of the above fields we must apply the
Laplace--Beltrami operator to the harmonic expansion.
We list such operators for the
$SO(5)$--harmonics\footnote{
Notice that the operator on the two--form $Y = Y_{ab}V^aV^b$ is
of the first order, like the fermionic ones.
Indeed it is the square root of the usual second order operator $\cD^a \cD_{[a}
Y_{bc]}$:
$$
\cD^a \cD_{[a} Y_{bc]} V^bV^c = \frac{1}{3} \star d \star d (Y_{ab}V^aV^b),
$$
where
$$
\star dY = \frac{1}{2} {\e_{ab}}^{cde} \cD_c Y_{de} V^a V^b.
$$
Hence $\frac{1}{2} {\e_{ab}}^{cde} \cD_c Y_{de} = \pm i \sqrt{3} \sqrt{\cD^c
\cD_{[c}} Y_{ab]}$.
} $Y_{[\l_1,\l_2]}^{(j,l)}$:
\bes
\boxtimes_y Y_{[0,0]} &\equiv& \Box Y, \\
\boxtimes_y Y_{[1,0]} &\equiv& 2 \cD^a \cD_{[a} Y_{b]}, \\
\boxtimes_y Y_{[1,1]} &\equiv& \star d Y_{ab} V^a V^b, \\
\boxtimes_y Y_{[2,0]} &\equiv& 3 \cD^c \cD_{(c}Y_{ab)}, \\
\boxtimes_y Y_{[1/2,1/2]} &\equiv& \cD\!\!\!\!\slash \;
\Xi, \\
\boxtimes_y Y_{[3/2,1/2]} &\equiv& \gamma^{abc} \, \cD_b \;
\Xi_c.
\ees

The explicit computation of the mass matrices derived from the above
Laplace--Beltrami differential operators will not be worked out here and we
refer the interested reader to \cite{CDD}.
We can give however in the simplest cases a couple of examples of the
computation.

\subsection{The scalar harmonic}

The case involving scalar harmonics $Y^{(j,l)}_{[0,0]}=Y_{q=0}^{j,l,r}$ is
straightforward.
In this case the five--dimensional invariant operator is simply the covariant
laplacian:
\beq
\Box = \cD^a\cD_a \equiv  \cD^i\cD_i + \cD^{s}\cD_{s} + \cD^5\cD_5.
\eeq

From \eqn{THv} and the fact that $T_{ab} L^{-1} \equiv T_{ab}
Y_{q=0}^{j,l,r} \equiv 0$, we obtain the following result
\beq
\label{box}
\Box Y_{q=0}^{j,l,r} = (-a^2 (T_i T_i + T_{s}T_{s}) - c^2 T_5 T_5)
Y_{q=0}^{j,l,r},
\eeq

Let us now evaluate \eqn{box}.
We set
\bea
T_i = -\frac{i}{2} \s_i, &&
T_{s} = -\frac{i}{2} \hat{\s}_{s}, \\
T_5 =T_3 - \hT_3 &=& \frac{i}{2} (\hat{\s}_3 - \s_3), \nonumber
\eea
where $\s$ and $\hat{\s}$ are ordinary Pauli matrices.
Using the relations
\bea
\s_1 \bet{|c|}\hline 1 \\\hline\eet =  \bet{|c|}\hline 2 \\\hline\eet
\qquad & \s_2 \bet{|c|}\hline 1 \\\hline\eet =  -i \bet{|c|}\hline 2 \\\hline\eet
 & \qquad \s_3 \bet{|c|}\hline 1 \\\hline\eet =  \bet{|c|}\hline 1 \\\hline\eet \\
\s_1 \bet{|c|}\hline 2 \\\hline\eet =  \bet{|c|}\hline 1 \\\hline\eet
\qquad & \s_2 \bet{|c|}\hline 2 \\\hline\eet =  i \bet{|c|}\hline 1 \\\hline\eet
 & \qquad \s_3 \bet{|c|}\hline 2 \\\hline\eet =  -\bet{|c|}\hline 2 \\\hline
\eet
\eea
(the same is true for $\hat{\s}$)
and observing that on a Young tableaux the $\s$'s act like a derivative
 (Leibnitz rule), we find on the first tableaux of \eqn{Tab2}
\bea
(\s_1\s_1 + \s_2\s_2)  \bet{|c|c|c|}\hline
 & \ldots & \\\hline
\eet &=& (2m_1(m_2+1) + 2m_2(m_1+1)) \bet{|c|c|c|}\hline
 & \ldots & \\\hline \eet = \\
&=& 4(j(j+1) - (j_3)^2)  \bet{|c|c|c|}\hline
 & \ldots & \\\hline \eet. \nonumber
\eea
An analogous result holds when acting with $\hat{\s}_1 \hat{\s}_1 + \hat{\s}_2
\hat{\s}_2$ on the second tableaux of \eqn{Tab2}, with $j\leftrightarrow l$.

Furthermore, the eigenvalue of $(\hat{\s}_3 - {\s}_3)^2$ on \eqn{Tab2} is
\beq
(m_2 - m_1 + n_2 -n_1)^2 = 4(j_3 + l_3)^2.
\eeq
For a scalar, $q=0$ and so, from \eqn{qr}, we have
\beq
j_3 = - l_3 = r/2.
\eeq
Therefore, we find
\beq
\Box Y_{(0)}^{(j,l,r)} = \left[a^2 j(j+1) +  b^2 l(l+1) +
(4c^2-a^2-b^2) \frac{r^2}{4}\right] Y_{(0)}^{(j,l,r)}.
\eeq
Substituting the values of $a$,$b$ and $c$ given  in \eqn{abc},
we obtain
\beq
\Box Y_{(0)}^{(j,l,r)} = H_0(j,l,r) Y_{(0)}^{(j,l,r)},
\eeq
where
\beq
\label{H0}
H_0(j,l,r)  \equiv 6\left(j(j+1) +  l(l+1) - \frac{r^2}{8}\right)
\eeq
is the eigenvalue of the Laplacian.
The same result was first given in \cite{G} using differential methods.

When the harmonic is not scalar, $q\neq0$,  the computation of the
Laplace Beltrami operators is more involved
since the covariant derivative \eqn{DaL-1} is valued
in the $SO(5)$ Lie algebra in the given representation $[\l_1,\l_2]$.

\subsection{The spinor harmonic}

We give as a further example the action of the $\cD\!\!\!\!\slash$ operator
on the spinor representation of $SO(5)$.
From \eqn{DaL-1} we have
\bea
\cD\!\!\!\!\slash &=& \gamma^a \cD_a = \gamma^i \left(-a T_i - \frac{a^2}{2c}
\e_{ij} T_5{}^j\right) +  \gamma^{s} \left(-a T_{s} + \frac{a^2}{2c}
\e_{st} T_5{}^{t}\right) +\nonumber \\
&+& \gamma^5 \left(-c T_5 -2 \left(c-\frac{a^2}{4c}
\right)(T_{12}-T_{34})\right),
\eea
where $T_{ab}$ are the $SO(5)$ generators in the spinor representation.
A straightforward computation gives
\beq
\label{matfer}
\cD\!\!\!\!\slash = \left(
\bet{cc}
$i c T_5 \unity_2 + \left(\dfrac{a^2}{4c} + c\right)\s^3$ & $-a \left(\s^i T_i
 + \s^3 \hT_1 - i \unity_2 \hT_2\right)$ \\
$a \left(\s^i T_i + \s^3 \hT_1 + i \unity_2 \hT_2\right)$ & $-ic T_5 \unity_2$
\eet
\right).
\eeq

When substituting the values of $c$ and $a$ in the matrix \eqn{matfer} we
note that \eqn{abc} defines them only up to a sign.
The right choice is dictated by supersymmetry.
Indeed, the existence of a complex Killing spinor $\eta(y)$ generating
$\cN = 2$ supersymmetry in $AdS_5$ implies that it must have the form
\beq
\label{eta}
\eta = \left(
\bet{c} $k$ \\ $l$ \\ 0 \\ 0
\eet
\right), \qquad k,l \in \hbox{\msbm C}
\eeq
since, being an $SU(2)\times SU(2)$ singlet, it must satisfy $T_H \eta = 0$
(see \eqn{THs}).
At this point the Killing equation
$\cD\!\!\!\!\slash \; \eta = \frac{5}{2} e \eta$ can be computed
from \eqn{matfer}  observing that on
an $SU(2) \times SU(2)$ singlet the $T_a$ generators have a null action
 and thus, using
\eqn{eta},
\beq
\cD\!\!\!\!\slash \; \eta = \left(
\bet{cc}
$\left(\dfrac{a^2}{4c} + c\right)\s^3$ & 0 \\
0 & 0
\eet
\right) \; \eta = \frac{5}{2} \, e \, \eta.
\eeq
This gives the correct value only if we choose $l=0$ and
\beq
c = - \frac{3}{2} e,
\eeq
while the sign of $a = \pm \sqrt{6} \, e$ is unessential.

Recalling the meaning of $c$ as the rescaling of the vielbein $V^5$,
{\it we conclude that $T^{11}$ admits a Killing spinor, leading to $\cN = 2$
supersymmetry on $AdS_5$, only for one orientation of $T^{11}$}.
To compute the mass matrix,
we write \eqn{matfer} as an explicit $4 \times 4$  matrix
\beq
\label{espli}
\cD\!\!\!\!\slash = e \left(
\bet{cccc}
$-i\dfrac{3}{2} T_5 + \frac{5}{2}$ & 0 & $\sqrt{6} \hT_+$ & $\sqrt{6} T_-$ \\
0 & $-i\dfrac{3}{2} T_5 - \frac{5}{2}$  & $\sqrt{6} T_+$ & $-\sqrt{6} \hT_-$ \\
$-\sqrt{6} \hT_-$ & $-\sqrt{6} T_-$ & $\dfrac{3}{2}i T_5$ & $0$ \\
$-\sqrt{6} T_+$ & $\sqrt{6} \hT_+$ & 0 & $\dfrac{3}{2}i T_5$
\eet
\right),
\eeq
where we have set
$$
T_\pm \equiv T_1 \pm i T_2, \qquad \hT_\pm \equiv \hT_1 \pm i \hT_2.
$$
$\cD\!\!\!\!\slash$ acts on the harmonic $\Xi = \left(\bet{c}
$Y_{(0)}$ \\
$Y_{(0)}$ \\
$Y_{(-1)}$ \\
$Y_{(+1)}$
\eet\right)^{(j,l)}
$
as a matrix whose entries are operators.

Since the harmonics are really defined up to a constant, the operatorial
matrix \eqn{espli} can be replaced by a numerical one, simply obtained by
substituting in each entry the values of the $T$--operators on the harmonics.
By diagonalization of this matrix one gets the eigenvalues which are related
to the fermion masses by numerical shifts. Analogous procedure can be used for
all the other invariant operators. In general the matrices can become very
large depending on the number of $U_H(1)$ fragments in the decomposition of
$[\l_1,\l_2]$. Leaving further explanations and all the details to the
forthcoming paper \cite{CDD}, we now quote the results for the mass spectrum.

\subsection{Spectrum and multiplet structure}

\begin{itemize}

\item
We begin by  the   spectrum   deriving from
the  {\bf scalar harmonic}  that appears in the expansion
of the ten--dimensional fields $h_{\mu\nu}(x,y)$, $B(x,y)$,
$h^{a}{}_{a}(x,y)$,
$A_{abcd}(x,y)$ and $A_{\mu\nu}$. The masses of the corresponding
five--dimensional fields (see table \eqn{fiel}) are thus given in terms of the
scalar harmonic eigenvalue  $H_0(j,l,r)$ given  in \eqn{H0}.
They are
\bea
\label{acca}
m^2(H_{\mu\nu}) &=& H_0,\\
m^2(B) &=& H_0,\\
\label{pibi}
m^2(\pi,b) &=& H_0 + 16 \pm 8 \sqrt{H_0+4},\\
m^2(a_{\mu\nu}) &=&  8 + H_0 \pm 4 \sqrt{H_0+4}\ .\label{amunu}
\eea
Note that while the laplacian acts diagonally on the $AdS_5$ fields
$H_{\mu\nu}(x)$ and $B(x)$, the  eigenvalues for $\pi(x)$ and $b(x)$,
which appear entangled in the linearised equations of motion \cite{KRV},
\cite{KW2}, have been obtained after diagonalisation of a two by two matrix.
With an abuse of notation,
in tables 2--10 we will call $\pi$, $b$ the linear combinations
given by the plus or minus signs in \eqn{pibi}.

\item
For the {\bf vector harmonic} we have found four eigenvalues
$$
\lambda_{[1,0]}=\{3 + H_0(j,l,r\pm 2),  H_0 + 4 \pm 2 \sqrt{H_0+4}\}.
$$
and the mass spectrum of the sixteen vectors is thus
\bea
m^2(a_{\mu}) &=& \left\{\begin{array}{c}
3 + H_0(j,l,r\pm 2) \\
 H_0 + 4 \pm 2 \sqrt{H_0+4} \end{array} \right. ,\\\label{bifi}
m^2(B_{\mu},\varphi_{\mu}) &=& \left\{\begin{array}{l}
H_0(j,l,r\pm 2) +7 \pm 4 \sqrt{H_0+4} \\
 H_0 + 12 \pm 6 \sqrt{H_0+4} \\
 H_0 + 4 \pm 2 \sqrt{H_0+4}
\end{array} \right.
\eea
In fact, as the Laplace--Beltrami operator acts diagonally on the complex vector
field $a_{\mu}(x)$ we get for it eight mass values .
Furthermore, the vectors $B_{\mu}(x)$, $\varphi_{\mu}(x)$ get mixed
in the linearised equations of motion, and upon diagonalisation we find
two extra masses for each eigenvalue. Here also we use the same names for
the linear combinations with plus or minus sign respectively
in the mass formulae \eqn{bifi} .

\item
For the {\bf antisymmetric tensor harmonics}
we get six eigenvalues from the Laplace Beltrami operator $\star d$
$$
\lambda_{[1,1]}=\left\{i\left(1\pm \sqrt{H_0(j,l,r\pm 2) + 4}\right),
\pm i \sqrt{H_0 +4}\right\}.
$$
and the masses
\bea
m^2(b_{\mu\nu})& =&\left\{
\begin{array}{l}
H_0+4\\
H_0+4\\
5+H_0(j,l,r\pm2)\pm 2\sqrt{H_0(j,l,r\pm2)+4}
\end{array}
\right. ,\\
m^2(a) &=& \left\{\begin{array}{l}
H_0 +4 \pm 4 \sqrt{H_0+4} \\
 H_0(j,l,r\pm 2) + 1 \pm 2 \sqrt{H_0(j,l,r\pm 2)+4}
\end{array} \right. .
\eea

\item
The {\bf spinor harmonics} eigenvalues of $\cD\!\!\!\!\slash$ 
are synthetically
$$
\lambda_{[\frac{1}{2},\frac{1}{2}]}=\left\{
\pm\frac{1}{2} \pm \sqrt{H_0(r\pm 1)+4}\right\}.
$$
The masses for the spinors and gravitinos are given in terms of
$\cD\!\!\!\!\slash$ by a numerical shift
\beq
\label{camp}
\begin{array}{rrcl}
\hbox{ gravitino : } & m(\psi_{\mu }) &=& \cD\!\!\!\!\slash -
\dfrac{5}{2}; \\
\hbox{ dilatino : } & m(\l) &=& \cD\!\!\!\!\slash +1; \\
\hbox{ longitudinal spinors: } & m(\psi^{(L)}) &=&
 \cD\!\!\!\!\slash +3;
\end{array}
\eeq
\end{itemize}
We have not yet calculated either the eigenvalues of $\cD\!\!\!\!\slash$
corresponding to the vector--spinor harmonic $\Xi_a$ which produce $AdS_5$
spinors
$\psi^{(T)}$, or the eigenvalues of the symmetric traceless harmonic
$Y_{(ab)}^{(\nu)}$.
However, we know a priori how many states we obtain in these two cases,
and by a counting argument we can circumvent the problem of the explicit
computation of the eigenvalues of
their mass matrices.
For the vector--spinors we have in principle a matrix of rank 20, that
becomes  $16\times 16$  due to the irreducibility condition, and further
gets to  $12\times 12$,  once the transversality condition
$\cD^a \Xi_a = 0$ is imposed.
In this way we are left with $12$  non--trivial (non longitudinal)
eigenvalues and thus we expect  $12$ $\psi^{(T)}$ spinors.
In an analogous way, the traceless symmetric tensor $Y^{(\nu)}_{(ab)}$
gives a $14\times 14$ mass--matrix  out of which five eigenvalues are
longitudinal leaving $9$ non--trivial eigenvalues .

If we match the bosonic and fermionic degrees of freedom
including the $12 + 12$ (right) left--handed spinors $\psi^{(T)}$ and
the $9$ real fields $\phi$ of the traceless symmetric tensor we find
$128$ bosonic degrees of freedom and $128$ fermionic ones.
Therefore, once we have correctly and unambiguously assigned all the fields
except the $\psi^{(T)}$ and $\phi$ to supermultiplets of $SU(2,2|1)$,
the remaining degrees of freedom of $\psi^{(T)}$ and $\phi$ are uniquely
assigned to the supermultiplets for their completion.

\bigskip

In tables 2--10 we have arranged our results in $SU(2,2|1)$
supermultiplets by an exaustion principle, starting from the highest
spin of the supermultiplet.
Each state of such multiplets is labelled by the $SU(2,2)$ quantum numbers
$(E_0,s_1,s_2)$ other than the internal symmetry attributes $(j,l,r)$.
As explained in section 3, $E_0$, the $AdS$ energy, is identified with the
conformal dimension $\Delta$.
Taking into account the $E_0$ value of each state and its $R$--symmetry,
we are able to fit unambiguously every mass at the proper place.
For this purpose it is essential to use the  relations between
the conformal weights $\Delta$ and the masses given by
\bea
\hbox{spin 2:} & \Delta = & 2 + \sqrt{4+m_{(2)}^2} \nonumber \\
\hbox{spin 3/2:} & \Delta = & 2 + |m_{(3/2)}+3/2| \nonumber \\
\label{Delta}
\hbox{spin 1:} & \Delta = & 2 + \sqrt{1+m_{(1)}^2}  \\
\hbox{two--form:} & \Delta = & 2 + |m_{(2f)}| \nonumber \\
\hbox{spin 1/2:} & \Delta_\pm = & 2 \pm |m_{(1/2)}| \nonumber \\
\hbox{spin 0:} & \Delta_{\pm} = & 2 \pm \sqrt{4+m_{(0)}^2} \nonumber
\eea
(where $\Delta$ is equal to the $E_0$ value of the state).
The sign ambiguity in the spin (0,$\frac{1}{2}$) dimensions is present because
the unitarity bound $E_0 \geq 1 +s$ allows the possibility $E_0 < 2$ for
such states.
The spin 0 case and its implications were analysed in \cite{KW2} and noticed
also in \cite{FGPW}.
There is no such ambiguities in all the other cases.

In the theory at hand, the chiral primary $Tr(AB)$ has the scalars with
$E_0 = \frac{3}{2}$, $E_0 + 1 = \frac{5}{2}$ coming from the $\Delta_{\pm}$
dimensions of the same $k=1$ mass value.
The fermionic partner is massless so there are no fermions with $E_0 < 2$.

\bigskip

We have found nine families of supermultiplets:
one graviton multiplet, four gravitino multiplets and four
vector multiplets which are reported in tables 2--10.

These are organised as follows.

In the first column we give the $(s_1,s_2)$ spin quantum numbers of the state.

In the second column we give the $E_0$ value of the state, where, according
to the standard nomenclature, the value of $E_0$ is referred to as the $E_0$ of
the multiplet and belongs to a vector field, a spin 1/2 field or to a scalar
field for the graviton, gravitino and vector multiplets respectively.
The other states have an $E_0$ value shifted in a range of $\pm 2$
(in 1/2 steps) with respect to the $E_0$ of the multiplet.

In the third column we write the $R$--symmetry of the state where the value
 $r$ is assigned to the highest spin state ($r = r^{h.s.}$),
the other states having $R$--symmetry shifted in a range of $\pm2$ (in
integer steps).

In the fourth column we give the right association of that particular
$SU(2,2|1)$ state to the field obtained from the KK spectrum, according
to the notations explained above.

In the fifth column we give the mass of the state\footnote{
Accordingly to \eqn{Delta} we give here the mass for the fermion and
two--form fields, while for all the other bosons we give the mass
squared.} in terms of the ubiquitous
expression $H_0$, where $H_0$ is evaluated  at a value $r$
corresponding to  that $R$--symmetry of the multiplet defined as
the $R$--symmetry of the highest spin $r = r^{h.s.}$.
We note that in all the formulae giving the mass spectrum
\eqn{acca}--\eqn{camp}, the $R$--symmetry $r$ refers to the
particular state
we are considering.
There, $H_0$ appears to have dependence on the $r$ of the state which is
different for different states.
However, when arranging the states in supermultiplets of $SU(2,2|1)$, it
is convenient to express the $r$ of the state in terms of the $R$--symmetry
of the supermultiplet $r=r^{h.s.}$,
defined as the $R$--symmetry of the highest spin.
In this case, all the masses can be expressed in terms of an $H_0$
 which has the same dependence on $r=r^{h.s.}$
for all the members of the multiplet.
For the graviton multiplet and the first two families of vector multiplets
all the masses are written in terms of $H_0 \equiv H_0(r)$;
for the (left) gravitino multiplets all the masses are given in terms of
$H_0^{\pm} \equiv H_0(j,l,r\pm 1)$ and for the last two families of
vectors all the masses are given in terms of $H_0^{\pm\pm} \equiv
H_0(j,l,r\pm 2)$.
Indeed, if we compute the conformal weight $\Delta$ of the state
from the mass values, it
turns out to be expressed in terms of $H_0$, $H_0^{\pm}$, $H_0^{\pm\pm}$
which are the same for every state of the multiplet, as it must be.
Of course, the value of $\Delta$ in terms of $H_0$, $H_0^{\pm}$, $H_0^{\pm\pm}$
 can be computed from \eqn{Delta}
and we have given for each multiplet the conformal weight of the lowest state
labelled by $E_0$ in terms of $H_0$.

The multiplets of  Tables 2--10 are long multiplets of
 $SU(2,2|1)$ when the $SU(2)\times SU(2)$ quantum numbers $j,l$ and the
$R$--symmetry values are generic. However,
it is well known from group theory \cite{FZ,FGPW} that 
shortening of the multiplets can occur in correspondence with particular values
of the  $SU(2,2|1)$ quantum numbers giving rise to chiral ($\bullet$),
semi--long ($\star$) or massless ($\diamond$) multiplets.
The above symbols have been used in the colums at the left of the tables to
denote the surviving states in the shortened multiplets.
In particular, the absence of these symbols in table 4 means that no
shortening of any kind can occur for the gravitino multiplet II.
Notice that shortenings are indicated only for positive values of the
 (shifted) $R$--symmetry $r$, namely when $r$ satisfies the following 
inequalities (see section 4)
\bea
r &\geq& 0 \ \hbox {Tables}\ 2,7,8\nonumber\\
r+1 &\geq& 0 \ \hbox{Tables}\ 4,5\nonumber\\
r-1 &\geq& 0  \ \hbox{Tables}\ 3,6\nonumber\\
r+2 &\geq& 0  \ \hbox{Table}\  9\nonumber\\
r-2 &\geq& 0  \ \hbox{Table}\  10.
\eea
In fact, these shortened multiplets
are the most interesting  in light of the
correspondence with the CFT at the boundary.
We give the discussion of the shortenings in section 4, after a preliminary
introduction to the representation of superconformal superfields in
CFT and the discussion of the conformal operators of protected
scaling dimensions.

\begin{table}[tbp]
\begin{center}
\caption{ {\bf Graviton Multiplet} \qquad $E_0=1+\sqrt{H_0+4}$.}
\label{ton}
\vskip .3cm
 \begin{tabular}{|c|c|c|c|c|c|c|}\hline
 & & $(s_1,s_2)$ &   $E_0^{(s)}$     &   $R$--symm.   &   field & Mass      \\
 \hline
 \hline
 $\diamond$ &$\star$ & (1,1)            & $E_0+1$      &$r$      &$H_{\mu\nu}$
     & $H_0$               \\\hline
  $\diamond$ &$\star$ &(1,{\small 1/2})          &$E_0+1/2$     &$r-1$    &$\psi_\mu^L$         &$-2+\sqrt{H_0+4}$     \\
  $\diamond$ &$\star$ &(1/2,1)          &$E_0+1/2$     &$r+1$    &$\psi_\mu^R$
         &$-2+\sqrt{H_0+4}$     \\
   &$\star$ &(1/2,1)          &$E_0+3/2$     &$r-1$    &$\psi_\mu^R$
     &$-2-\sqrt{H_0+4}$   \\
   & &(1,1/2)          &$E_0+3/2$     &$r+1$    &$\psi_\mu^L$
&$-2-\sqrt{H_0+4}$      \\
 \hline
  $\diamond$ &$\star$ &(1/2,1/2)     &$E_0$  &$r$    &$ \phi_\mu$ &$H_0+4-2\sqrt{H_0+4}$    \\
   & &(1/2,1/2)     &$E_0+1$&$r+2$  &$ a_\mu$  &$H_0+3$  \\
   &$\star$ &(1/2,1/2)     &$E_0+1$&$r-2$  &$ a_\mu$   & $H_0+3$  \\
   & &(1/2,1/2)     &$E_0+2$&$r$    &$B_\mu$   &$H_0+4+2\sqrt{H_0+4}$   \\
\hline
   & &(1,0)     & $E_0+1$      &$r$  &$b_{\mu\nu}^+$ &$\sqrt{H_0+4}$    \\
   &$\star$ &(0,1)     & $E_0+1$      &$r$  &$b_{\mu\nu}^-$ &$-\sqrt{H_0+4}$     \\
\hline
   & &(1/2,0) & $E_0+1/2$      &$r+1$  & $\lambda_L$&$1/2-\sqrt{H_0+4}$  \\
   &$\star$ &( 0,1/2)& $E_0+1/2$      &$r-1$  & $\lambda_R$&$1/2-\sqrt{H_0+4}$   \\
   & &(1/2,0)  & $E_0+3/2$     &$r-1$  & $\lambda_L$&$1/2+\sqrt{H_0+4}$    \\
   & &(0,1/2)  & $E_0+3/2$ &$r+1$  &$\lambda_R$  &$1/2+\sqrt{H_0+4}$  \\
\hline
   & &(0,0)  & $E_0+1$&$r$ & $ B$ & $H_0$     \\
\hline
 \end{tabular}
 \end{center}
\end{table}


\begin{table}[tbp]
\begin{center}
\caption{ {\bf Gravitino Multiplet I}\qquad $E_0=\sqrt{H_0^-+4}-1/2$}
\label{tin}
\vskip .3cm
 \begin{tabular}{|c|c|c|c|c|c|c|}\hline
 & & $(s_1,s_2)$ &   $E_0^{(s)}$     &   $R$--symm.   &   field & Mass      \\
 \hline \hline
&$\star$ & (1,1/2) &$E_0+1$&$r$&$\psi_\mu^L$&$-3+\sqrt{H_0^-+4}$\\
 \hline
   &$\star$ &(1/2,1/2)          &$E_0+1/2$     &$r+1$    &$\phi_\mu$         &$H_0^-+7-4\sqrt{H_0^-+4}$  \\
   &$\star$ &(1/2,1/2)          &$E_0+3/2$     &$r-1$    &$ a_\mu$         &$H_0^-+4-2\sqrt{H_0^-+4}$     \\
 \hline
  $\bullet$ &$\star$ &(1,0)          &$E_0+1/2$     &$r-1$    &$ a_{\mu\nu}$
  &$2-\sqrt{H_0^-+4}$  \\
   & &(1,0)        &$E_0+3/2$     &$r+1$    &$b_{\mu\nu}^+$         &$1-\sqrt{H_0^-+4}$      \\
 \hline
  $\bullet$ &$\star$ &(1/2,0)     &$E_0$  &$r$    &$ \psi^{(T)}_L$ &$-5/2+\sqrt{H_0^-+4}$    \\
   $\bullet$&$\star$ &(1/2,0)     &$E_0+1$&$r-2$  &$\psi^{(T)}_L$  &$-3/2+\sqrt{H_0^-+4}$  \\
   &$\star$ &(0,1/2)     &$E_0+1$&$r$  &$\lambda_R$   & $3/2-\sqrt{H_0^-+4}$  \\
   & &(1/2,0)     &$E_0+1$&$r+2$    &$\psi^{(T)}_L$   &$-3/2+\sqrt{H_0^-+4}$   \\
   & &(1/2,0)     &$E_0+2$      &$r$  &$\psi^{(T)}_L$ &$-1/2+\sqrt{H_0^-+4}$    \\
 \hline
   $\bullet$&$\star$ &(0,0)     & $E_0+1/2$      &$r-1$  &$a$ &$H_0^-+4-4\sqrt{H_0^-+4}$    \\
   & &(0,0)     & $E_0+3/2$      &$r+1$  & $a$&$H_0^-+1-2\sqrt{H_0^-+4}$  \\
\hline
 \end{tabular}
 \end{center}
\end{table}

\begin{table}[tbp]
\begin{center}
\caption{{\bf Gravitino Multiplet II}\qquad $E_0=5/2 +\sqrt{H_0^++4}$
 }
\label{tinnon}
\vskip .3cm
 \begin{tabular}{|c|c|c|c|c|}\hline
   $(s_1,s_2)$ &   $E_0^{(s)}$     &   $R$--symm.   &   field & Mass      \\
 \hline
 \hline
(1,1/2)&$E_0+1$&$r$&$\psi_\mu^L$&$-3-\sqrt{H_0^++4}$\\
 \hline
(1/2,1/2)          &$E_0+1/2$     &$r+1$    &$a_\mu$         &$H_0^++4+2\sqrt{H_0^++4}$     \\
(1/2,1/2)          &$E_0+3/2$     &$r-1$    &$B_\mu$         &$H_0^++7+4\sqrt{H_0^++4}$\\
 \hline
(1,0)          &$E_0+1/2$     &$r-1$    &$b_{\mu\nu}^+$         &$1+\sqrt{H_0^++4}$      \\
(1,0)        &$E_0+3/2$     &$r+1$   &$a_{\mu\nu}$         &$2+\sqrt{H_0^++4}$  \\
\hline
(1/2,0)     &$E_0$  &$r$    &$ \psi^{(T)}_L$ &$-1/2-\sqrt{H_0^++4}$    \\
(1/2,0)     &$E_0+1$&$r-2$  &$\psi^{(T)}_L$  &$-3/2-\sqrt{H_0^++4}$  \\
(0,1/2)     &$E_0+1$&$r$  &$\lambda_R$   & $3/2+\sqrt{H_0^++4}$  \\
(1/2,0)     &$E_0+1$&$r+2$    &$\psi^{(T)}_L$   &$-3/2-\sqrt{H_0^++4}$   \\
(1/2,0)     &$E_0+2$      &$r$  &$\psi^{(T)}_L$ &$-5/2-\sqrt{H_0^++4}$    \\
 \hline
(0,0)     & $E_0+1/2$      &$r-1$  &$a$ &$H_0^++1+2\sqrt{H_0^++4}$    \\
(0,0)     & $E_0+3/2$      &$r+1$  & $a$&$H_0^++4+4\sqrt{H_0^++4}$  \\
\hline
 \end{tabular}
 \end{center}
\end{table}


\begin{table}[tbp]
\begin{center}
\caption{ {\bf Gravitino Multiplet III}\qquad $E_0=-1/2+\sqrt{H_0^++4}$}
\label{tin2}
\vskip .3cm
 \begin{tabular}{|c|c|c|c|c|c|}\hline
  & $(s_1,s_2)$ &   $E_0^{(s)}$     &   $R$--symm.   &   field & Mass      \\
 \hline \hline
$\star$ & (1/2,1) &$E_0+1$&$r$&$\psi_\mu^R$&$-3+\sqrt{H_0^++4}$\\
 \hline
   $\star$ &(1/2,1/2)          &$E_0+1/2$     &$r-1$    &$\phi_\mu$         &$H_0^++7-4\sqrt{H_0^++4}$  \\
    &(1/2,1/2)          &$E_0+3/2$     &$r+1$    &$ a_\mu$         &$H_0^++4-2\sqrt{H_0^++4}$     \\
 \hline
  $\star$ &(0,1)          &$E_0+1/2$     &$r+1$    &$ a_{\mu\nu}$         &$2-\sqrt{H_0^++4}$  \\
   $\star$ &(0,1)        &$E_0+3/2$     &$r-1$    &$b_{\mu\nu}^-$         &$1-\sqrt{H_0^++4}$      \\
 \hline
  $\star$ &(0,1/2)     &$E_0$  &$r$    &$ \psi^{(T)}_R$ &$-5/2+\sqrt{H_0^++4}$    \\
   $\star$ &(0,1/2)     &$E_0+1$&$r+2$  &$\psi^{(T)}_R$  &$-3/2+\sqrt{H_0^++4}$  \\
    &(1/2,0)     &$E_0+1$&$r$  &$\lambda_L$   & $3/2-\sqrt{H_0^++4}$  \\
    &(0,1/2)     &$E_0+1$&$r-2$    &$\psi^{(T)}_R$   &$-3/2+\sqrt{H_0^++4}$   \\
    &(0,1/2)     &$E_0+2$      &$r$  &$\psi^{(T)}_R$ &$-1/2+\sqrt{H_0^++4}$    \\
 \hline
    &(0,0)     & $E_0+1/2$      &$r+1$  &$a$ &$H_0^++4-4\sqrt{H_0^++4}$    \\
    &(0,0)     & $E_0+3/2$      &$r-1$  & $a$&$H_0^++1-2\sqrt{H_0^++4}$  \\
\hline
 \end{tabular}
 \end{center}
\end{table}

\begin{table}[tbp]
\begin{center}
\caption{{\bf Gravitino Multiplet IV}\qquad $E_0=5/2+\sqrt{H_0^-+4}$
 }
\label{tinnon2}
\vskip .3cm
 \begin{tabular}{|c|c|c|c|c|c|}\hline
  & $(s_1,s_2)$ &   $E_0^{(s)}$     &   $R$--symm.   &   field & Mass      \\
 \hline
 \hline
$\star$&(1/2,1)&$E_0+1$&$r$&$\psi_\mu^R$&$-3-\sqrt{H_0^-+4}$\\
 \hline
$\star$&(1/2,1/2)          &$E_0+1/2$     &$r-1$    &$a_\mu$         &$H_0^-+4+2\sqrt{H_0^-+4}$     \\
&(1/2,1/2)          &$E_0+3/2$     &$r+1$    &$B_\mu$         &$H_0^-+7+4\sqrt{H_0^-+4}$\\
 \hline
$\star$&(0,1)          &$E_0+1/2$     &$r+1$    &$b_{\mu\nu}^-$         &$1+\sqrt{H_0^-+4}$      \\
$\star$&(0,1)        &$E_0+3/2$     &$r-1$   &$a_{\mu\nu}$         &$2+\sqrt{H_0^-+4}$  \\
 \hline
$\star$&(0,1/2)     &$E_0$  &$r$    &$ \psi^{(T)}_R$ &$-1/2-\sqrt{H_0^-+4}$    \\
$\star$&(0,1/2)     &$E_0+1$&$r+2$  &$\psi^{(T)}_R$  &$-3/2-\sqrt{H_0^-+4}$  \\
&(1/2,0)     &$E_0+1$&$r$  &$\lambda_L$   & $3/2+\sqrt{H_0^-+4}$  \\
&(0,1/2)     &$E_0+1$&$r-2$    &$\psi^{(T)}_R$   &$-3/2-\sqrt{H_0^-+4}$   \\
&(0,1/2)     &$E_0+2$      &$r$  &$\psi^{(T)}_R$ &$-5/2-\sqrt{H_0^-+4}$    \\
 \hline
&(0,0)     & $E_0+1/2$      &$r+1$  &$a$ &$H_0^-+1+2\sqrt{H_0^-+4}$    \\
&(0,0)     & $E_0+3/2$      &$r-1$  & $a$&$H_0^-+4+4\sqrt{H_0^-+4}$  \\
\hline
 \end{tabular}
 \end{center}
\end{table}


\begin{table}[tbp]
\begin{center}
\caption{{\bf Vector Multiplet I} \qquad $E_0=\sqrt{H_0+4}-2$ }
\label{tor1}
\vskip .3cm
 \begin{tabular}{|c|c|c|c|c|c|c|c|}\hline
& &    & $(s_1,s_2)$    & $E_0^{(s)}$ &   $R$--symm.   &   field & Mass      \\
 \hline
 \hline
$\diamond$&&$\star$&(1/2,1/2)            &$E_0+1$      &$r$
&$\phi_\mu$&$H_0+12-6\sqrt{H_0+4}$\\
 \hline
$\diamond$& $ \bullet$  &$\star$ &(1/2,0)          &$E_0+1/2$     &$r-1$    &$\psi^{(L)}_L$         &$7/2-\sqrt{H_0+4}$  \\
$\diamond$&   &$\star$  &(0,1/2)                  &$E_0+1/2$     &$r+1$    &$\psi^{(L)}_R$         &$7/2-\sqrt{H_0+4}$     \\
  & &$\star$  &(0,1/2)          &$E_0+3/2$     &$r-1$    &$\psi^{(L)}_R$         &$5/2-\sqrt{H_0+4}$  \\
  & & &(1/2,0)&$E_0+3/2$     &$r+1$    &$\psi^{(L)}_L$         &$5/2-\sqrt{H_0+4}$      \\
 \hline
 $\diamond$& $\bullet$ &$\star$ &(0,0)     &$E_0$  &$r$    &$b$ &$H_0+16-8\sqrt{H_0+4}$    \\
  & $\bullet$&$\star$ &(0,0)     &$E_0+1$&$r-2$  &$\phi$  &$H_0+9-6\sqrt{H_0+4}$  \\
  & & &(0,0)     &$E_0+1$&$r+2$  &$\phi$   & $H_0+9-6\sqrt{H_0+4}$  \\
  & & &(0,0)     &$E_0+2$&$r$    &$\phi$   &$H_0+4-4\sqrt{H_0+4}$   \\
 \hline
 \end{tabular}
 \end{center}
\end{table}

\begin{table}[tbp]
\begin{center}
\caption{{\bf Vector Multiplet II} \qquad $E_0=\sqrt{H_0+4}+4$ }
\label{tor2}
\vskip .3cm
 \begin{tabular}{|c|c|c|c|c|}\hline
      $(s_1,s_2)$   & $E_0^{(s)}$ &   $R$--symm.   &   field & Mass      \\
 \hline
 \hline
 (1/2,1/2)            &$E_0+1$      &$r $&$B_\mu  $&$H_0+12+6\sqrt{H_0+4}$\\
 \hline
   (1/2,0)        &$E_0+1/2$     &$r-1$    &$\psi^{(L)}_L$         &$5/2+\sqrt{H_0+4}$  \\
   (0,1/2)        &$E_0+1/2$     &$r+1$    &$\psi^{(L)}_R$         &$5/2+\sqrt{H_0+4}$     \\
  (0,1/2)        &$E_0+3/2$     &$r-1$    &$\psi^{(L)}_R$         &$7/2+\sqrt{H_0+4}$  \\
    (1/2,0)        &$E_0+3/2$     &$r+1$    &$\psi^{(L)}_L$         &$7/2+\sqrt{H_0+4}$      \\
 \hline
  (0,0)     &$E_0$  &$r$    &$\phi$ &$H_0+4+4\sqrt{H_0+4}$    \\
   (0,0)     &$E_0+1$&$r-2$  &$\phi$  &$H_0+9+6\sqrt{H_0+4}$  \\
    (0,0)     &$E_0+1$&$r+2$  &$\phi$   & $H_0+9+6\sqrt{H_0+4}$  \\
    (0,0)     &$E_0+2$&$r$    &$\pi$   &$H_0+16+8\sqrt{H_0+4}$   \\
 \hline
 \end{tabular}
 \end{center}
\end{table}

\begin{table}[tbp]
\begin{center}
\caption{{\bf Vector Multiplet III} \qquad
$E_0=\sqrt{H_0^{++}+4}+1$;
 }
\label{tor3}
\vskip .3cm
 \begin{tabular}{|c|c|c|c|c|c|}\hline
     & $(s_1,s_2)$    & $E_0^{(s)}$ &   $R$--symm.   &   field & Mass      \\
 \hline
 \hline
&(1/2,1/2)            &$E_0+1$      &$r$       &$ a_\mu$&$H_0^{++}+3$\\
 \hline
    &(1/2,0)          &$E_0+1/2$     &$r-1$    &$\psi^{(T)}_L$        &$-1/2+\sqrt{H_0^{++}+4}$  \\
    &(0,1/2)                  &$E_0+1/2$     &$r+1$    &$\psi^{(T)}_R$         &$-1/2+\sqrt{H_0^{++}+4}$     \\
    &(0,1/2)          &$E_0+3/2$     &$r-1$    &$\psi^{(T)}_R$         &$1/2+\sqrt{H_0^{++}+4}$  \\
   $\bullet$ &(1/2,0)        &$E_0+3/2$     &$r+1$    &$\psi^{(T)}_L$         &$1/2+\sqrt{H_0^{++}+4}$      \\
 \hline
   &(0,0)     &$E_0$  &$r$    &$a$ &$H_0^{++}+1-2\sqrt{H_0^{++}+4}$    \\
    &(0,0)     &$E_0+1$&$r-2$  &$\phi$  &$H_0^{++}$  \\
   $\bullet$ &(0,0)     &$E_0+1$&$r+2$  &$\phi$   & $H_0^{++}$  \\
   $\bullet$ &(0,0)     &$E_0+2$&$r$    &$a$   &$H_0^{++}+1+2\sqrt{H_0^{++}+4}$   \\
 \hline
 \end{tabular}
 \end{center}
\end{table}
\begin{table}[tbp]
\begin{center}
\caption{{\bf Vector Multiplet IV} \qquad
$E_0=\sqrt{H_0^{--}+4}+1$ }
\label{tor4}
\vskip .3cm
 \begin{tabular}{|c|c|c|c|c|c|c|}\hline
&     & $(s_1,s_2)$   & $E_0^{(s)}$ &   $R$--symm.   &   field & Mass      \\
 \hline
 \hline
&$\star$ &(1/2,1/2)            &$E_0+1$      &$r$               &$a_\mu$&$H_0^{--}+3$\\
 \hline
  $\bullet$ &$\star$ &(1/2,0)       &$E_0+1/2$     &$r-1$   &$\psi^{(T)}_L$          &$-1/2+\sqrt{H_0^{--}+4}$  \\
   &$\star$ &(0,1/2)       &$E_0+1/2$     &$r+1$    &$\psi^{(T)}_R$               &$-1/2+\sqrt{H_0^{--}+4}$     \\
   &$\star$ &(0,1/2)       &$E_0+3/2$     &$r-1$   &$\psi^{(T)}_R$          &$1/2+\sqrt{H_0^{--}+4}$  \\
   & &(1/2,0)       &$E_0+3/2$     &$r+1$    &$\psi^{(T)}_L$          &$1/2+\sqrt{H_0^{--}+4}$      \\
 \hline
 $\bullet$  &$\star$ &(0,0)     &$E_0$  &$r$    &$a$ &$H_0^{--}+1-2\sqrt{H_0^{--}+4}$    \\
 $\bullet$ &$\star$  &(0,0)     &$E_0+1$&$r-2$  &$B$  &$H_0^{--}$  \\
   & &(0,0)     &$E_0+1$&$r+2$  &$\phi$   & $H_0^{--}$  \\
   & &(0,0)     &$E_0+2$&$r$    &$a$   &$H_0^{--}+1+2\sqrt{H_0^{--}+4}$   \\
 \hline
 \end{tabular}
 \end{center}
\end{table}

\newpage

\section{CFT and  $SU(2,2|1)$ representations}

\subsection{$SU(2,2|1)$ conformal superfields}

The $AdS/CFT$ correspondence \cite{M,GKP,W} gives a relation between
the particle states in $AdS_5$, classified in this case by the
$SU(2,2|1)$ superalgebra and the realisation of the very same
representations \cite{GKP,W,FF} in terms of conformal fields on the
boundary $\tilde{M}_4 = \partial AdS_5$.

In this way, the highest weight representations of $SU(2,2|1)$ correspond
to {\it primary} superconformal fields on the boundary and a generic state
on the bulk, labelled by four quantum numbers \cite{FZ,FF1,DP} $\cD (E_0,
s_1, s_2 | r)$ related to $U(1) \times SU(2) \times SU(2) \times
U_R(1) \subset SU(2,2) \times U_R(1)$, is mapped to a primary conformal
field $\cO_{(s_1, s_2)}^{\Delta,r}(x)$ with scaling dimension $\Delta=E_0$,
Lorentz quantum numbers $(s_1,s_2)$ and $R$-symmetry $r$.
$E_0$ is the $AdS$ energy level and its relation to the $AdS$ mass
depends on the spin of the state.
We recall here the relevant cases \cite{W,FZ,FFZ}
\beq
\bet{crcl}
$\left(\frac{1}{2},\frac{1}{2}\right)$ & $m^2$  &=& $(E_0 - 1) (E_0 - 3) $ \\
$(0,0)$ & $m^2$ &=& $E_0 (E_0 - 4) $ \\
$(1,0)$, $(0,1)$ & $m^2$ &=& $(E_0 - 2)^2 $ \\
$(1,1)$ & $m^2$ &=& $E_0 (E_0 - 4)  $ \\
$\left.
\bet{c}
$(\frac{1}{2},0)$, $(0,\frac{1}{2}),$\\
$(\frac{1}{2},1)$, $(1,\frac{1}{2})$
\eet
\right\}$
 & $m$ &=& $|E_0 - 2| $.
\eet
\eeq

It is crucial in our discussion to classify states corresponding to
short multiplets because in this case the conformal dimension $\Delta$ is
{\it protected} and it allows a stringent test between the
supergravity theory and the conformal field theory realisation.
Here, protected means that $\Delta$ is related to the $R$--charge
which is quantised in terms of the isometry generator of $U_R(1)$.
However, we note that unlike the $\cN = 4$ theory \cite{GKH,DS}, operators with
protected dimensions have conformal dimension different from their
free--field value.

$\cN = 1$ superfields with protected and unprotected dimensions
have been discussed by many authors \cite{W,FZ,FGPW,O}.
We would like to remind here just their field theory realisation, which will
become especially important in comparing conformal operators with the
particular model described by the IIB theory compactified on $AdS_5
\times T^{11}$.

A generic conformal primary superfield is classified by an
$SL(2,\IC)$ $(s_1,s_2)$
representation, a dimension $E_0$ and an $R$--symmetry charge $r$.
These are the quantum numbers of the $\th = 0$ component of the
superfield.
All descendants are given by the $\th$ expansion which also dictates
their spin, $R$--symmetry $r$ and scaling dimension $\Delta$,
 since $\th_{\a}$ has
$(s_1,s_2) = (\frac{1}{2},0)$, $\Delta=-\frac{1}{2}$, $r = 1$
(so $\bth_{\da}$ has
$(s_1,s_2) = (0,\frac{1}{2})$, $\Delta=-\frac{1}{2}$, $r = -1$).
For a generic primary conformal field the dimension is not
protected since it can take any value $\Delta\geq 2 + s_1 + s_2$ ($s_1 s_2
\neq 0$) or $\Delta \geq 1 + s$ ($s_1 s_2 = 0$) due to unitarity bounds of
the irrepses of $SU(2,2)$ \cite{BFH}.
$SU(2,2|1)$ requires the additional unitarity bounds 
\beq
\label{addi}
2+2s_1 - E_0
\leq \frac{3}{2} r \leq E_0 -2 -2s_2,
\eeq
$E_0 \geq 1+s$ ($E_0 =\frac{3}{2}|r|$),
$E_0 = s_1 = s_2 = r = 0$ (identity representation),
which restrict the allowed values of the $R$--symmetry charge 
\cite{FGPW,FF1,DP}.

\bigskip

Operators with protected dimensions fall in four categories (as
discussed in \cite{FZ,FGPW,O})
\begin{enumerate}
    \item {\bf Chiral superfields}: $S$
    They satisfy the condition
    \beq
    \bar{D}_{\da} S_{(\a_1 \ldots \a_{2s_1})}(x,\th,\bth) = 0.
    \eeq
    For them $s_2 = 0$ ($s_1 = 0$ if antichiral) and $r = \dfrac{2}{3} \Delta$
    $\left(r = - \dfrac{2}{3} \Delta \hbox{ if antichiral}\right)$.
    These superfields contain the (massless on the boundary) free 
    {\it singleton} representations for $\Delta=1+s$.
    These multiplets have $4(2s+1)$ degrees of freedom.
    \item {\bf Semichiral superfields}: $U_{\a_1 \ldots
    \a_{2s_1}, \da_1 \ldots \da_{2s_2}}$
    They satisfy the condition
    \beq
    \bar{D}_{(\da} U_{\da_1 \ldots \da_{2s_2}), \a_1 \ldots
    \a_{2s_1}}(x,\th,\bth) = 0,
    \eeq
    and for them $ r = \dfrac{2}{3} (\Delta+2 s_2)$.
    If  $s_2 = 0$ the above superfield becomes chiral.
    For example $s_2 = \frac{1}{2}$ would correspond to
    semichiral superfield whose lowest component is a right--handed spin $1/2$
    and its highest spin is a vector field with $r = \dfrac{2}{3} \Delta-\frac{1}{3}$.
    \item {\bf Conserved superfields}: $J_{(s_1,s_2)}$
    They satisfy
    \beq
    D^{\a_1} J_{\a_1 \ldots \a_{2s_1}, \da_1 \ldots \da_{2s_2}}
    (x,\th,\bth) = 0,
    \eeq
    and
    \beq
    \bar{D}^{\da_1} J_{\a_1 \ldots \a_{2s_1}, \da_1 \ldots \da_{2s_2}}
    (x,\th,\bth) = 0,
    \eeq
    (or $\bar{D}^2 J_{\a_1 \ldots \a_{2s_1}} = 0$ if $s_2 = 0$) and
    for them $r = \dfrac{2}{3} (s_1 - s_2)$, $\Delta = 2 + s_1 + s_2$.
    \item {\bf Semi--conserved superfields}: $L_{(s_{1}, s_2)}$
    They satisfy
    \beq
    \bar{D}^{\da_1} L_{\a_1 \ldots \a_{2s_1}, \da_1 \ldots \da_{2s_2}}
    (x,\th,\bth) = 0,
    \eeq
    or
    \beq
    \bar{D}^{2} L_{\a_1 \ldots \a_{2s_1}}
    (x,\th,\bth) = 0 \qquad \hbox{ for  }  s_2 = 0.
    \eeq
    Their $R$--symmetry is $r = \dfrac{2}{3} (\Delta-2-2s_2)$.
    A semi--conserved superfield becomes conserved if it is left and
    right semi--conserved in which case
    $\Delta = 2+s_1+s_2$ and $r = \dfrac{2}{3} (s_1-s_2)$.
\end{enumerate}

Operators of type 1) 2) and 4) have protected (but anomalous) dimensions in a
non--trivial conformal field theory.
They are short or semishort because some of the fields in the $\th$
expansion are missing.
In the language of \cite{FGPW} the 1) and 2) superfields
correspond to the shortening conditions $n_2^+ = 0$ ($n_1^+=0$),
3) correspond to
$n_1^- = n_2^- = 0$ and 4) to $n_2^- = 0$ ($n_1^-=0$).

In the $AdS/CFT$ correspondence all these superfields
correspond to KK states
with multiplet shortening and typically they occur when there is a
lowering in the rank of the mass
matrix and rational values of $E_0$ are obtained.
Conserved current multiplets correspond to massless fields in $AdS_5$.
They can only occur for fields whose mass is protected by a symmetry
(such as gauge fields) and there is only a finite number of them
corresponding to the gauge fields of the $SU(2,2|1)\times SU(2) \times SU(2)$
algebra and possibly Betti multiplets \cite{Betti,Betti2}.
While the massless vectors of the isometry group correspond to the
$U_R(1)$ and flavour symmetry of the boundary gauge theory, the Betti
multiplet, as recently shown by Klebanov and Witten \cite{KW2},
corresponds to the $U_b(1)$ baryonic current multiplet of the boundary CFT.
There are also two complex moduli related to $B$ and $A_{ab}$ wrapped 
on a 2--cycle of $T^{11}$ \cite{KW}, giving two hypermultiplets with $E_{0} = 
3$ and $r = 2$.
Massive KK states with arbitrary irrational value of $E_0$
correspond to generic conformal field operators with anomalous
dimension.

It is easy to relate operators of different type by superfield
multiplication.
By multiplying a chiral $(s_1,0)$ by an anti--chiral $(0,s_2)$
primary one gets a generic superfield with $(s_1,s_2)$, $\Delta = \Delta^c +
\Delta^a$ and $r = \frac{2}{3}(\Delta^c - \Delta^a)$.
By multiplying a {\it conserved current} superfield $J_{\a_1 \ldots
\a_{2s_1}, \da_1 \ldots \da_{2s_2}}$ by a chiral scalar
superfield one gets a semi--conserved superfield with $\Delta =
\Delta^{c} + 2 + s_1 + s_2$ ($r = \frac{2}{3}(\Delta-2-2s_2)$).

In a KK theory only particular values of $(s_1,s_2)$ can occur,
because the theory in higher dimensions has only spin 2, spin 3/2
fields and lower.
This implies that for bosons only $(0,0)$, $(1,0)$, $(0,1)$,
$\left(\frac{1}{2},\frac{1}{2}\right)$, $(1,1)$ representations and
for fermions only  $\left(\frac{1}{2},0\right)$, $\left(0,\frac{1}{2}\right)$,
$\left(1,\frac{1}{2}\right)$, $\left(\frac{1}{2},1\right)$ representations can occur.
This drastically limits the spin of conformal superfields.
Indeed, for chiral ones $s=0,\frac{1}{2}$, while for non chiral $s_1, s_2
\leq \frac{1}{2}$.

\subsection{CFT analysis of $AdS_5 \times T^{11}$ compactification}

In the conformal field theory \cite{KW} which is {\it dual} to IIB
supergravity on $AdS_5 \times T^{11}$ the basic superfields are the
gauge fields\footnote{Below we use standard superfield notations \cite{BW}.}
 $W_{\a}$ of $SU(N)\times SU(N)$ and two doublets of chiral
superfields $A$, $B$ which are in the $(N,\bar{N})$ and $(\bar{N},N)$
of $SU(N)\times SU(N)$ and in the $\left(\frac{1}{2},0\right)$ $r=1$,
($0,\frac{1}{2}$) $r=1$
of the {\it global} symmetry group $SU(2) \times SU(2) \times U_R(1)$.
At the conformal point these superfields have anomalous dimension
$\Delta = 3/4$ and $R$--symmetry $r=1/2$.
The chiral $W_\alpha$ superfield has $\Delta=3/2$, $r=1$.

Let us specify the superspace gauge transformations of the above superfields.
Following \cite{BW}, we introduce Lie algebra valued chiral parameters
$\Lambda_1$, $\Lambda_2$ of the two factors of $\cG=SU(N)\times SU(N)$.
Then, under $\cG$ gauge transformations
\beq
\begin{array}{rcl}
e^{V_1} & \longrightarrow & e^{i \Lambda_1} e^{V_1} e^{-i
\bar\Lambda_1} \\
e^{V_2} & \longrightarrow  & e^{i \Lambda_2} e^{V_2} e^{-i
\bar\Lambda_2} \\
A & \longrightarrow  & e^{i \Lambda_1} A e^{-i \Lambda_2} \\
B & \longrightarrow  & e^{i \Lambda_2} B e^{-i \Lambda_1}
\end{array}
\eeq
and we define
\beq
\begin{array}{rcl}
W_{1\a} &=& \bar D\bar D \left( e^{V_1} D_\a e^{-V_1}\right) \\
W_{2\a} &=& \bar D\bar D \left( e^{V_2} D_\a e^{-V_2} \right)
\end{array}
\eeq
where $V_1$ and $V_2$ are superfields Lie algebra valued in the two $\cG$
 factors and $V=V_1+V_2$.
Gauge covariant combinations are therefore
\bea
W_\a (AB)^k &=&W_\a^1 (AB)^k\label{w1}\\
W_\a (BA)^k&=&W_\a^2 (BA)^k\label{w2}\\
A e^V \bar A e^{-V} &=&A e^{V_2} \bar A e^{-V_1}\label{a1}\\
B e^V \bar B e^{-V} &=&B e^{V_1} \bar B e^{-V_2}\label{bb1}
\eea
Formulae \eqn{w1} and \eqn{a1} transform as
\beq
X\longrightarrow e^{i\Lambda_1} X e^{-i \Lambda_1}
\eeq
while \eqn{w2} and \eqn{bb1} transform as
\beq
Y\longrightarrow e^{i\Lambda_2} Y e^{-i \Lambda_2}
\eeq
We can multiply \eqn{a1} and \eqn{bb1} as
\beq
A e^{V_2} \bar A \bar B e^{-V_2} B \label{c1}
\eeq
which transforms as $X$ or
\beq
B e^{V_1} \bar B \bar A e^{-V_1} A \label{c2}
\eeq
which transforms as $Y$ and thus build gauge covariant combinations as
$W^1_\a X$ or $W^2_\a Y$.

If a symmetry $A\leftrightarrow B$ is required, then symmetrization
exchanging \eqn{w1} with \eqn{w2}, \eqn{a1} with \eqn{bb1} or
\eqn{c1} with \eqn{c2} will occur.

We will now consider sets of towers of superfields, labelled by an
integer number $k$ which correspond to {\it chiral} and {\it (semi--)conserved}
gauge invariant superfields and having therefore protected dimensions.
As we will see in the next section,
these conformal operators are precisely those corresponding to $AdS$--KK
states undergoing multiplet shortening.

\bigskip

Let us first consider chiral superfields.
There are three infinite sequences of them, corresponding to
{\it hypermultiplets} and {\it tensor multiplets} in the $AdS$ bulk.

They are given as\footnote{Here and in what follows we always 
mean symmetrized trace and symmetrized
$SU(2)\times SU(2)$ indices.}:
\bea
\label{Sk}
S^k = Tr(AB)^k, & \Delta^k = \frac{3}{2} k, & r = k, \quad k > 0,\\
\label{Tk}
T^k = Tr\,\left(W_{\a}(AB)^k\right), & \Delta^k = \frac{3}{2} (k+1), &
r = k+1, \quad k >0,\\
\label{Fik}
\Phi^k = Tr\,\left(W^{\a}W_{\a}(AB)^k\right), & \Delta^k = 3 + \frac{3}{2} k,
& r = k+2.
\eea

The series \eqn{Sk} was anticipated by Klebanov, Witten \cite{KW} and
shown to occur in the KK modes of the supergravity theory by Gubser
\cite{G}, who also discussed descendants of the series \eqn{Fik}.

The series \eqn{Tk}--\eqn{Fik} have been constructed by the knowledge
of the full mass spectrum and the shortening conditions\footnote{
Chiral operators of the type $Tr(W_{\a_1} \ldots W_{\a_p})$ cannot appear
in the KK spectrum for $p > 2 $ since such operators have 
$ \Delta = \frac{3}{2}p$,
$r=p$, $j=l=0$ and therefore are incompatible with the spectrum 
of the $U_R(1)$ charge
on $T^{11}$ (see next section).
For $p=2$ the chiral operators 
$Tr(W_{\a_1} W_{\a_2} (AB)^k)$ are allowed but they
contain two irreducible parts: one symmetric ($(1,0)$ spin one) and the other
antisymmetric ($(0,0)$ spin zero).
However, following an observation of Aharony
 (as quoted in \cite{VAFA}) only the scalar  term
is a chiral primary operator.
This is due to the superspace identity
$$
\bar{D} \bar{D} \left[ e^V D_{\a} 
\left( e^{-V} W_{\b} e^V \right) e^{-V} \right]
= \left[ W_{\a}, W_{\b}\right],
$$
where the symmetry of the left hand side derives 
from the following superspace Bianchi identity
$
e^V D^{\a}\left( e^{-V} W_{\a} e^V \right) e^{-V} =
\bar D_{\da}\left( e^{V} \bar W^{\da} e^{-V} \right).
$
Therefore, the other term is not chiral primary since
$$
Tr(W_{(\a} W_{\b)} (AB)^k) = \bar D \bar D Tr\left( e^V D_{\a} \left(
e^{-V} W_{\b} e^V \right) e^{-V} (AB)^k \right).
$$
}.

It is useful to note that in the \eqn{Tk} and \eqn{Fik}
towers, we find operators of the type
\bea
\label{Bmunu}
B_{\a\b}^k = Tr(F_{\a\b}(AB)^k), & & \Delta^k = 2 + \frac{3}{2} k, \; (k>0)\\
\label{fik}
\phi^k = Tr(F_{\a\b}F^{\a\b}(AB)^k), & & \Delta^k = 4 + \frac{3}{2} k,
\eea
as descendants.  $F_{\a\b}$, $F_{\da\db}$ refer in the
spinor notation to the dual and anti--selfdual parts of the field strength
$F_{\mu\nu}$.

\bigskip

Even more interesting is the appearance of (semi--)conserved superfields
corresponding in the language of \cite{FGPW} to semilong multiplets
in $AdS_5$.
These superfields explain the appearance of KK towers with  (spin
1) vector fields and (spin 2) tensor fields with protected dimensions.

In superfield language such fields are given by superfields
containing terms of the form
\bea
\label{9}
{J}_{\a\da}^k = Tr({J}_{\a\da}(AB)^k), &&
\left\{
\bet{ll}
$j=l=\dfrac{k}{2}$,  & $r=k$,\\
$\Delta=3+\dfrac{3}{2} k$ & \eet \right. \\
\label{10}
{J}^k = Tr({J}(AB)^k), &&  \left\{
\bet{ll}
$j=l+1$,  $l=\dfrac{k}{2}$, & $r=k$,  \\
$\Delta=2+\dfrac{3}{2}k$  &
\eet
\right. \\
\label{Ik}
{I}^k = Tr({J}W^2 (AB)^k) && \left\{
\bet{ll}
$j=l+1$,  $l=\dfrac{k}{2}$, &
$r=k+2$, \\
$\Delta=5+\dfrac{3}{2} k$ &
\eet
\right.
\eea
where
\bea
\label{11}
{J}_{\a\da} &=& W_{\a} e^V \bar W_{\da} e^{-V},  \qquad (\Delta=3),\\
\label{12}
{J}  &=& A (e^V \bar{A}) e^{-V},   \qquad (\Delta=2),
\eea
and satisfying
\beq
\bar{D}^{\da}{J}_{\a\da}^k = 0, \qquad \bar{D}\bar{D}{J}^k = 0,
\qquad \bar{D}\bar{D}{I}^k=0.
\eeq
Analogous structures appear with $B$ replacing $A$ in \eqn{12}
and $j\leftrightarrow l$ in \eqn{10} and \eqn{Ik}.
Note that the {\it non gauge invariant} operators in \eqn{9}--\eqn{Ik}
behave as if they would
have conformal dimension 3 and 2 respectively when the gauge singlet
is formed.
This is because the shortening condition implies that operators
starting with structures as in \eqn{9}, \eqn{10} and \eqn{Ik}
have dimension given by $3 + \frac{3}{2} k$ and $2 + \frac{3}{2}
k$ and $5+\frac{3}{2}k$ respectively.

The highest spin states contained in \eqn{9}, \eqn{10} and \eqn{Ik} are {\it
descendants} with spin 2  and $\Delta = 4 + \frac{3}{2} k$, spin
1 with $\Delta = 3+ \frac{3}{2} k$ and spin 1 with $\Delta =6+\frac{3}{2}k$.
These are massive recursions of the graviton, massless  gauge boson and
massive vector fields respectively.
The $AdS$ masses of the above states are given by
\bea
\label{massgrav}
\displaystyle
\hbox{spin 2:} && M^k = \sqrt{\frac{3}{2} k \left( \frac{3}{2} k +
4\right)}, \\
\label{massvec}
\displaystyle
\hbox{spin 1:} && M^k = \sqrt{\frac{3}{2} k \left( \frac{3}{2} k +
2\right)}, \\
\label{massvec2}
\displaystyle
\hbox{spin 1:} && M^k = \sqrt{\left(\frac{3}{2} k+5 \right)
\left(\frac{3}{2} k+3\right)}.
\eea
The first two masses vanish for the $k = 0$ level corresponding to the {\it
conserved} currents $Tr {J}_{\a\da}$, $Tr {J}$ of the superconformal field
theory with flavour group $G=SU(2)\times SU(2)$,
while the third mass does not vanish at $k=0$.

For the spin 3/2 massive tower  we do not expect to get vanishing gravitino mass
when $k=0$, since the massless gravitino is already contained in the
graviton tower.
In spite of this, there are semi--conserved superfields corresponding
to shortened massive gravitino towers.

These are
\bea
\label{elleda1}
L^{1k}_{\da} &=& Tr\left(e^V \bar{W}_{\da} e^{-V} (AB)^k\right),
\left\{ \begin{array}{rcl} j&=&l,\quad  r=k-1,\\
\Delta&=&\frac{3}{2}+\frac{3}{2} k \end{array}  \right. \ k > 0,\\
\label{2Lda}
L^{2k}_{\da} &=& Tr\left(e^V \bar{W}_{\da} e^{-V} W^2(AB)^k\right),
\left\{ \begin{array}{rcl}
j&=&l,\quad r=k+1,\\ \Delta&=&\frac{9}{2}+\frac{3}{2} k \end{array}  \right.\\
&\hbox{ and }&\nonumber \\
\label{La}
L^{3k}_{\a} &=& Tr\left(W_{\a}\, (A e^V \bar{A} e^{-V})\, (AB)^k\right),
\left\{ \begin{array}{rcl}  j&=&l+1,\quad r=k+1,\\ \Delta&=&\frac{7}{2}+
\frac{3}{2} k \end{array}  \right.
\eea
which satisfy $\bar{D}^{\da} L_{\da} =0$ and $\bar{D}^2 L_{\a}
= 0$, respectively.

We note in particular that the tower analogous to \eqn{elleda1}, in type IIB
supergravity on $AdS_5 \times S^5$ is \cite{W,FFZ,AF,marzaf}
\beq
L^{1k}_{\da} = Tr( e^V \bar W_{\da} e^{-V} \phi_{(i_1} \ldots \phi_{i_k)})
\eeq
in the $k$--fold symmetric of $SU(3)$.
For $k>1$ these superfields are semiconserved but for $k=1$, unlike
in our case, they become conserved, corresponding to the fact that on
$S^5$ an additional $SU(3)$ triplet of massless gravitinos is
required by $\cN = 4$ supersymmetry.

In this case the exact operator $L^{11}_{\da}$ is
\beq
L^{11}_{\da} = Tr \left[ (e^{V} \bar{W}_{\da} e^{-V} \phi_a) +
\bar{D}_{\da}(e^V \bar{\phi}^b e^{-V}) \, (e^V \bar{\phi}^c e^{-V})
\; \e_{abc}\right]
\eeq
which satisfies
\beq
\bar{D}^{\da}L^{11}_{\da} = {D}^{2}L^{11}_{\da} =0
\eeq
as a consequence of the equations of motion for $W_{\a}$, $\phi_a$ and
the identity
\beq
D^2 \left[ e^{-V} \bar{D}_{\da}(e^{V} \bar{\phi}^a e^{-V})
e^V\right] =
 \left[ \bar{\phi}^a ,\bar{W}_{\da} \right].
\eeq

The above superfields \eqn{elleda1}--\eqn{La} are the lowest 
non--chiral operators of
more general towers with irrational scaling dimensions described by
\bea
O^{1nk}_{\da} &=& Tr\left(e^V \bar{W}_{\da} e^{-V} (A e^V \bar{A}
e^{-V})^n (AB)^k\right), \\
O^{2nk}_{\da} &=& Tr\left(e^V \bar{W}_{\da} e^{-V} (A e^V \bar{A}
e^{-V})^n W^2 (AB)^k\right), \hbox{ and }\\
O^{3nk}_{\a} &=& Tr\left(W_{\a}\, (A e^V \bar{A} e^{-V})^n\, (AB)^k\right),
\eea
with $G$ representation
\bea
O^{1nk}_{\da}: & \left(\frac{k}{2}+n,\frac{k}{2}\right), & r=k-1, \\
O^{2nk}_{\da}: &\left(\frac{k}{2}+n,\frac{k}{2}\right), & r=k+1, \\
O^{3nk}_{\a}: &\left(\frac{k}{2}+n,\frac{k}{2}\right), & r=k+1.
\eea

The multiplets in \eqn{Sk}--\eqn{Fik}, \eqn{9}--\eqn{Ik} and \eqn{elleda1}--\eqn{La} are
shortened multiplets with protected dimensions
because of supersymmetry through non--renormalisation theorems.
However we will see that a peculiar phenomenon of $\cN=1$
which can be learned from the
$AdS/CFT$ correspondence is that there exist also infinite towers of long
multiplets with rational dimensions which in principle are not expected to have
protected dimensions .

A typical tower which is not expected to have protected dimension is
the massive tower
\beq
\label{nonpro}
Q^k=Tr (\; W^2 e^V \bar{W}^2 e^{-V} (AB)^{k})
\eeq
which contains the descendant
$Tr(F_{\alpha\beta}F^{\alpha\beta} \bar F_{\dot\alpha\dot\beta}\bar 
F^{\dot\alpha\dot\beta} (AB)^k)$.
Supergravity predicts for it $\Delta=8+\frac{3}{2} k$

We just note that the analogous operator in type IIB on $AdS_5 \times
S^5$ was a descendant of a chiral primary (showing up at first at
$p=4$ level \cite{GM,FFZ,AF,marzaf}) and therefore having protected
dimensions because of $\cN = 4$ supersymmetry \cite{GKH,DS,KTV}.

The identification of such long  multiplets with superconformal operators
will be given in the next section.
Operators whose $R$--symmetry is not related to the top components of
one of the two $SU(2)$ factors (see section 4) are for instance towers of the form
\beq
Tr \left[ ( Ae^V \bar A e^{-V})^{n_1}(e^V \bar B e^{-V}  B )^{n_2}(AB)^k\right],
\eeq
which have $j=\frac{k}{2}+n_1$, $l=\frac{k}{2}+n_2$ and $r=k$.
These operators have all irrational dimensions unless $n_1$, $n_2$ are
consecutive terms in a particular sequence described in \cite{G}.

\medskip

It is worthwhile to point out
 that in this gauge theory we have
no realisation of the semi--chiral superfields described before and
indeed we do not find on the supergravity side any shortened multiplet
satisfying the
$r=\frac{2}{3}(E_0+2 s_2)$ condition ($s_2 \neq 0$).
The reason is that such superfield  correspond to 
non--unitary modules.

\section{AdS/CFT correspondence}

In section 2 and 3 we have described the KK spectrum with its multiplet structure and
the CFT operators with protected dimensions.
We would like now to
present the multiplet shortening conditions and analyse the correspondence
of these states with the boundary field theory operators shown in the last
section.
This is an important non--trivial check for the $AdS/CFT$ correspondence.
On the other hand, supergravity seems to suggest additional
dynamical inputs to the extent that it predicts that certain towers of long
multiplets have rational dimensions, suggesting the presence of some
hidden symmetry.
This latter may perhaps be explained in the context of Born--Infeld theory
which relates $D$--brane dynamics to $AdS$ supergravity in the large $N$
limit.

From the point of view of the $SU(2,2|1)$ multiplet structure, the shortening
conditions correspond to saturation of some of the inequalities describing
the unitarity bounds \cite{FGPW}.
These become relations between $E_0$ and the other $SU(2,2|1)$ quantum numbers.

In the KK context, we do not  know a priori the
multiplet structure of the KK states and the shortening conditions merely
derive from the disappearance of some harmonics in the field expansion.
This reduces the rank of the mass matrices and thus some of the
states drop from the multiplet.
The relevant fact is that these shortening conditions must be in one to one
correspondence with those deriving from the $SU(2,2|1)$ group theoretical
analysis.

As discussed in the previous section, the shortening conditions can be read
as the following relations on the $SU(2,2|1)$ quantum numbers
already given in section (3.1)
\bea
\label{b1}
\hbox{ (anti--) chiral } && E_0= + \frac{3}{2}r \; \left(-\frac{3}{2}r\right),\\
\label{b2}
\hbox{ conserved } && E_0= 2 + s_1 + s_2, \qquad  (s_1 - s_2) =\frac{3}{2}r, \\
\label{b3}
\hbox{ semi--conserved }&& E_0 = \frac{3}{2}r +2 s_{2}+2, \qquad (\hbox{or }
s_{2} \to  s_1, r\to -r).
\eea

This means that the corresponding conformal dimension must have a rational
value.
As it can easily be seen from the mass spectrum presented in section two,
this implies that only for specific $G$ quantum numbers we can retrieve
such short multiplets.
Actually, a rational scaling dimension can be found only if $H_0(j,l,r) +4$
is a perfect square of a rational number.
Two possible sets of values for which such a condition is satisfied are:
\bea
\label{AA}
&j=l=\left|\dfrac{r}{2}\right|&=\dfrac{k}{2} \\
\label{BB}
j = l- 1 = \left|\dfrac{r}{2}\right| &=\dfrac{k}{2} \hbox{or} 
& l=j-1 =\left|\dfrac{r}{2}\right|=\dfrac{k}{2}
\eea
We will also examine briefly the case 
\beq
\label{CC}
j = l = \frac{r-2}{2} \qquad r \geq 2,
\eeq
which for most multiplets leads to a violation of inequality \eqn{addi}, 
but in one case  gives a consistent shortening of the vector multiplet III. 
We will show that these three cases are the relevant ones.
Indeed, in the first case $H_0(j,l,r) = \dfrac{9}{4}r^2 + 6|r|$ and thus
$H_0(j,l,r)+4=\left(3\left|\dfrac{r}{2}\right|+2\right)^2$,
in the second $H_0(j,l,r) = \dfrac{9}{4}r^2 + 12|r| + 12$ and thus
$H_0(j,l,r)+4=\left(3\left|\dfrac{r}{2}\right|+4\right)^2$,
 while in the third case 
we have $H_0(j,l,r) = \dfrac{9}{4}r^2 - 6r$ and thus
$H_0(j,l,r)+4=\left(3\dfrac{r}{2}-2\right)^2$.

Of course there are other possible solutions, but we will see that only those
presented above correspond to multiplet shortening.

Looking at the tables 2--10 we see that for the graviton and type I and II
vector multiplets (V.M.) $E_0$ is given in terms of $H_0(j,l,r)$
while for  gravitino multiplet of type I, IV and II, III $E_0$ is given in
terms of $H_0^{\mp} \equiv H_0(j,l,r\mp 1)$ respectively.
Analogously, for the type III and IV  V.M., $E_0$ is given in terms of
$H_0^{\pm\pm} \equiv H_0(j,l,r\pm 2)$ respectively.
As a consequence the conditions for rational values of $E_0$ (protected
dimensions) are different for different multiplets.

Let us examine the conditions \eqn{AA},\eqn{BB} and \eqn{CC} separately.

Condition \eqn{AA} for the various multiplets reads
\bea
\label{gra}
\hbox{Graviton and type I and II V.M. } && j= l = \left|\frac{r}{2}\right|
\equiv \frac{k}{2}, \\
\label{grv1}
\hbox{type I gravitino } && j= l = \left|\frac{r- 1}{2}\right|\equiv
\frac{k}{2}, \\
\label{grv2}
\hbox{type II gravitino } && j= l = \left|\frac{r+ 1}{2}\right|\equiv
\frac{k}{2}, \\
\label{grv3}
\hbox{type III gravitino } && j= l = \left|\frac{r+ 1}{2}\right|\equiv
\frac{k}{2}, \\
\label{grv4}
\hbox{type IV gravitino } && j= l = \left|\frac{r- 1}{2}\right|\equiv
\frac{k}{2}, \\
\label{ve3}\hbox{type III V.M. } && j= l = \left|\frac{r+2}{2}\right|\equiv
\frac{k}{2}, \\
\label{ve4}\hbox{type IV V.M. } && j= l = \left|\frac{r-2}{2}\right|\equiv
\frac{k}{2},
\eea
Here $k\in \hbox{\msbm Z}_+$ identifies the $SU(2)\times SU(2)$ representations
of the multiplet; it is obvious that all the multiplets obeying condition
\eqn{AA} are in the irrep ($\frac{k}{2}$,$\frac{k}{2}$).

Substituting in the $E_0$ value of the multiplet given in tables 2--10
$H_0+4$, $H_0^{\pm}+4$ and $H_0^{\pm\pm} + 4$
with $\left(\frac{3}{2}k + 2\right)^2$ we find the
following values of $E_0$ for the various multiplets
\bea
\label{bgra}
\hbox{Graviton multiplet } & E_0 = \dfrac{3}{2}k + 3 & \equiv 
\pm \frac{3}{2}r + 3,
 \\\label{bgrv1}
\hbox{type I L.H. gravitino  } & E_0 = \dfrac{3}{2}k + \frac{3}{2} &
\equiv \left\{ 
\bet{l} $\dfrac{3}{2}r$\\ $-\dfrac{3}{2}r + 3$\eet \right., \\\label{bgrv2}
\hbox{type II L.H. gravitino  } & E_0 = \dfrac{3}{2}k + \frac{9}{2} &
\equiv \left\{ \bet{l} $\dfrac{3}{2}r + 6$ \\ $-\dfrac{3}{2}r+3$\eet \right., \\\label{bgrv3}
\hbox{type III R.H. gravitino   } & E_0 = \dfrac{3}{2}k + \frac{3}{2} &
\equiv \left\{ \bet{l} $\dfrac{3}{2}r + 3$ \\ $-\dfrac{3}{2}r$\eet \right., \\\label{bgrv4}
\hbox{type IV R.H. gravitino   } & E_0 = \dfrac{3}{2}k + \frac{9}{2} &
\equiv \left\{ \bet{l} $\dfrac{3}{2}r + 3$ \\ $-\dfrac{3}{2}r+6$\eet \right.,
 \\\label{bve1}
\hbox{type I V.M. } & E_0 = \dfrac{3}{2}k & \equiv \pm \frac{3}{2}r , 
\\\label{bve2}
\hbox{type II V.M. } & E_0 = \dfrac{3}{2}k + 6 & \equiv \pm \frac{3}{2}r + 6, \\
\label{bve3}
\hbox{type III V.M. } & E_0 = \dfrac{3}{2}k + 3 & \equiv \left\{ \bet{l} 
$\dfrac{3}{2}r + 6$ 
\\ $-\dfrac{3}{2}r$\eet \right., \\
\label{bve4}
\hbox{type IV V.M. } & E_0 = \dfrac{3}{2}k + 3 & \equiv \left\{ \bet{l} $\dfrac{3}{2}r$
\\ $-\dfrac{3}{2}r+6$\eet \right., 
\eea
where the upper and lower choices on the right hand side refer to positive or 
negative  arguments 
of the absolute values in \eqn{gra}--\eqn{ve4}.

Using \eqn{b1}--\eqn{b3} we see that under condition \eqn{AA}
we obtain:
\begin{itemize}
\item
a chiral tensor multiplet from type I L.H. gravitino \eqn{bgrv1}
 (or an antichiral one from type III R.H. gravitino);
\item
one hypermultiplet (for both signs of $r$) from type I
 V.M \eqn{bve1}, and
 another hypermultiplet from type IV V.M.  \eqn{bve4} (or from type III V.M
if $r<-2$) ;
\item
a semilong graviton multiplet from \eqn{bgra} (for both signs of $r$);\\
two semilong gravitino from type III and IV (or from type I  if $r<1$ and
type II if $r<-1$ respectively), and IV  R.H. gravitino multiplets
from  the two equations \eqn{bgrv3} and \eqn{bgrv4}.

\item
For $k=0$ ($G$--singlet),  we also obtain from \eqn{bgra} a short
massless graviton multiplet with $E_0=3$, $r=0$.
In this case only four states
survive: the massless graviton, two massless gravitini (with $r=\pm1$ depending
on the chirality), and one massless vector.
This latter, being an $SU(2)\times SU(2) \times U_R(1)$ singlet, must be 
identified with the $R$--symmetry Killing vector.

\end{itemize}
Note that \eqn{bgrv2}, \eqn{bve2} and \eqn{bve3} do not correspond to any
shortening
condition, yet we have a rational value of $E_0$ belonging to a long
multiplet.

It is now easy to find the correspondence between the supermultiplets obeying
condition \eqn{gra}--\eqn{ve4}
 and the primary conformal superfields on the CFT side
discussed in the previous section.
Given the values of $E_0$ and $k$ (or $r$) we have immediately
that the two hypermultiplets from \eqn{bve1} and \eqn{bve4} are in 
correspondence
with the chiral superfields $S^k$ and $\Phi^k$
 \eqn{Sk} and \eqn{Fik};
the tensor multiplet from \eqn{bgrv1} corresponds to the chiral
superfield $T^k$
of \eqn{Tk};
the semilong graviton multiplet from \eqn{bgra}, associated with
 the semi--conserved
superfield $J_{\a\da}^k$ of \eqn{9}
(in particular the massless graviton multiplet ($k=0$ in \eqn{bgra})
corresponds to
the conserved superfield $J_{\a\da}^0$);
finally, the two semilong gravitino multiplets from \eqn{bgrv3} and \eqn{bgrv4}
can be put in correspondence with the semi--conserved superfields
$L_{\da}^{1,k}$ and $L_{\da}^{2k}$
of \eqn{elleda1} and \eqn{2Lda}.

We note that  the type I vector series in Table 7 for $j=l=r=0$,
see \eqn{bve1}, degenerates into the identity representation, since $E_0 = 0$.
However, as follows from the same table, another unitary representation, a
massless vector multiplet, appears in the spectrum. 
Indeed, for $j=l=r=0$,
the multiplet bosonic mass squared eigenvalues 
are $m_{(1)}^2 = 0$, $m_{(0)}^2 = 0$, $m_{(0)}^2 = -3$, $m_{(0)}^2 = -4$.
The eigenvalue $m_{(0)}^2 = 0$ gives two possible values for $E_0$:
$E_0 = 0$ and $E_0 = 4$.
If we choose the $E_0 = 0 $ branch, the other modes
(scalars with $E_0 = 1,2$ and vector with $E_0 = 1$)  are gauge modes
and decouple from the physical Hilbert space, thus the multiplet is a
gauge module \cite{BFH}.
If we choose the $E_0 = 4 $ branch, we get a unitary
representation with a scalar with $E_0 =2 $ and a vector with $E_0 = 3$ as
physical states, while the other modes (scalars with $E_0= 3,4$) decouple from
the physical Hilbert space.
This massless vector multiplet is the so called {\it Betti multiplet} of KK
supergravity, related to the fact that a $(p+1)$--form (in this case
$p=3$) couples to a $p$--brane wrapped on a non--trivial $p$--cycle 
which in this case is
related to $b_3 = 1$, the third Betti number of $T^{11}$ \cite{KW2,GKcan}.
The general occurrence of such Betti multiplets in the KK context was
widely discussed in \cite{Betti}.
In the case of $AdS_4 \times M^{111}$, such a multiplet is related to 
$b_2 = 1$
\cite{Betti2,Wittold}, corresponding to the $M$--theory three--form with
one component on $AdS_4$ and two components on $M^{111}$ and it was found
in the KK context in \cite{M111}.
Incidentally, in the language of \cite{FF86}, the Betti massless
vector ($D(3,1/2,1/2)$ ) is a zero center module\footnote{
A zero center module also appears in the graviton multiplet of the
$OSp(6|4)$ superalgebra \cite{FF86}. In fact this multiplet contains
an $O(6)$ singlet massless vector other than the $O(6)$ gauge
fields. 
This agrees with the geometrical interpretation of $\cN = 6$ supergravity as 
the low--energy limit of type $IIA$ string theory on $AdS_4 \times 
\hbox{\msbm CP}^3$, the latter being obtained by Hopf reducing $M$--theory on 
$AdS_4 \times S^7$ \cite{N6}.}
of the conformal group $SU(2,2)$,
since all the Casimir vanish $C_I = C_{II} = C_{III} = 0$  as
is  the case for the  identity
$D(0,0,0)$,the gauge module $D(1,1/2,1/2)$, the massless scalars 
$D(4,0,0)$ appearing in the hypermultiplet $S^k$ for $k=0$ \eqn{Sk} 
 and the spin one singleton $D(2,1,0) + D(2,0,1)$
representations \cite{BFH,FF86,fefr}.
The geometrical origin of this gauge field coupled to a wrapped D3 brane 
on $T^{11}$
has recently been discussed in \cite{KW2} together with its
interpretation as baryon current in the $AdS/CFT$ correspondence.

The boundary superfield corresponding to the Betti multiplet is
\beq
\label{cU}
{\cal U}=Tr\ A e^V \bar A e^{-V} - Tr\ B e^V \bar B e^{-V} \qquad 
(D^2 {\cal U}=\bar D^2 {\cal U}=0).
\eeq
Its $\th =0$ component is a scalar $\left.{\cal U}\right|_{\th = 0}
= A \bar A - B \bar B$ with $E_0 = 2$ ($m^2_{(0)}=-4$) and
the baryon current is the
$\theta\sigma_\mu \bar\theta$ component with
 $\Delta=E_0+1=3$ ($m^2_{(1)}=0$) \cite{KW2}. 
Note that all KK states are neutral under
the $U_B(1)$, and thus it lies outside the $T^{11}$ isometry.

Beside shortened multiplets, there are CFT superconformal operators with
rational dimensions that are associated with  the long multiplets of
\eqn{bgrv2},\eqn{bve2}
 and \eqn{bve3}.
Indeed we may construct the following superfields\footnote{
The $Q^k$ massive tower was also considered in \cite{G}.} all in the
$(\frac{k}{2},\frac{k}{2})$ of $G$:
\bea
P^k_{\a} &=& Tr\left(W_{\a} e^{V} \bar{W}^2 e^{-V} (AB)^k\right)
\qquad \Delta = \frac{3}{2}k+\frac{9}{2}, \quad r=k-1, \; k>0, \\
Q^k &=& Tr\left(W^2 e^{V} \bar{W}^2 e^{-V} (AB)^k\right)
\qquad \Delta = \frac{3}{2}k+6, \quad r=k, \\
R^k &=& Tr\left(e^{V} \bar{W}^2 e^{-V} (AB)^k\right)
\qquad \Delta = \frac{3}{2}k+3, \quad r=k-2, \; k>0.
\eea
Let us now discuss the shortening conditions when the $G$--quantum numbers
satisfy condition \eqn{BB}.

In this case \eqn{gra}--\eqn{ve4} are replaced by the analogous equations
\bea
\label{cgra}
\hbox{Graviton and type I and II V.M. } && l= j-1
= \left|\frac{r}{2}\right|\equiv
\frac{k}{2}, \\\label{cgrv1}
\hbox{type I gravitino } && l= j- 1 = \left|\frac{r- 1}{2}\right|\equiv
\frac{k}{2}, \\\label{cgrv2}
\hbox{type II gravitino } && l= j - 1 = \left|\frac{r+ 1}{2}\right|\equiv
\frac{k}{2}, \\\label{cgrv3}
\hbox{type III gravitino } && l= j- 1 = \left|\frac{r+ 1}{2}\right|\equiv
\frac{k}{2}, \\\label{cgrv4}
\hbox{type IV gravitino } && l= j - 1 = \left|\frac{r- 1}{2}\right|\equiv
\frac{k}{2}, \\\label{cve3}
\hbox{type III V.M. } && l= j - 1= \left|\frac{r+2}{2}\right|\equiv
\frac{k}{2}, \\\label{cve4}
\hbox{type IV V.M. } && l= j - 1= \left|\frac{r-2}{2}\right|\equiv
\frac{k}{2},
\eea
(or $j \leftrightarrow l$)
where all the states have the representation ($\frac{k}{2}+1$,$\frac{k}{2}$)
if $j=l+1$ or in the  ($\frac{k}{2}$,$\frac{k}{2}+1$)
if $l=j+1$.

Proceeding as before we now substitute $H_0+4$, $H_0^{\pm}+4$, $H_0^{\pm\pm}+4$
with $\left(
\frac{3}{2}k+4\right)^2$ in the $E_0$--value of the various multiplets given
in tables 2--10 and we obtain for each multiplet the following rational
values of $E_0$:
\bea\label{dgra}
\hbox{Graviton multiplet } & E_0 = \dfrac{3}{2}k + 5 & \equiv \frac{3}{2}r + 5,
 \\\label{dgrv1}
\hbox{type I L.H. gravitino  } & E_0 = \dfrac{3}{2}k + \frac{7}{2} &
\equiv \frac{3}{2}r+2, \\\label{dgrv2}
\hbox{type II L.H. gravitino  } & E_0 = \dfrac{3}{2}k + \frac{13}{2} &
\equiv \frac{3}{2}r + 8, \\\label{dgrv3}
\hbox{type III R.H. gravitino  } & E_0 = \dfrac{3}{2}k + \frac{7}{2} &
\equiv \frac{3}{2}r+5, \\\label{dgrv4}
\hbox{type IV R.H. gravitino  } & E_0 = \dfrac{3}{2}k + \frac{13}{2} &
\equiv \frac{3}{2}r + 5, \\\label{dve1}
\hbox{type I V.M. } & E_0 = \dfrac{3}{2}k +2& \equiv \frac{3}{2}r +2, \\
\label{dve2}
\hbox{type II V.M. } & E_0 = \dfrac{3}{2}k + 8 & \equiv \frac{3}{2}r + 8, \\
\label{dve3}
\hbox{type III V.M. } & E_0 = \dfrac{3}{2}k + 5 & \equiv \frac{3}{2}r + 8, \\
\label{dve4}
\hbox{type IV V.M. } & E_0 = \dfrac{3}{2}k + 5 & \equiv \frac{3}{2}r+2,
\eea
where we have limited ourselves to the positive branch of the 
expressions in the absolute values appearing in \eqn{cgrv1}--\eqn{cve4}.

By \eqn{b1} we see that there are no chiral
supermultiplets when condition \eqn{BB} holds.
However we have that \eqn{dgrv1}, \eqn{dve1} and \eqn{dve4} give the
condition \eqn{b3} for
semilong multiplets, all the other values of $E_0$ corresponding
to long multiplets with rational dimensions.

Thus we have:
one semilong type I L.H. gravitino corresponding to the semi--conserved
superfield \eqn{La};
one semilong type I V.M. corresponding to the semi--conserved superfield
$J^k$ of \eqn{10} which, in  the particular case $k=0$, becomes a conserved
superfield $J$ corresponding to the massless type I V.M. with $E_0 = 2$,
$r=0$ (these correspond to the $SU(2) \times SU(2)$ Killing vectors);
one semilong type IV V.M. corresponding to the semi--conserved
superfield $I^k$ of \eqn{Ik}.

Furthermore we have long multiplets from \eqn{dgra}, \eqn{dgrv2},
\eqn{dgrv3}, \eqn{dgrv4},
\eqn{dve2},\eqn{dve3} corresponding respectively to the following
superconformal fields with rational dimensions
\bea
C^k &=& Tr\left( A e^V \bar{A} e^{-V} J_{\a\da} (AB)^k\right), \qquad
E_0 = \frac{3}{2}k + 5, \quad r = k, \\
D^k &=& Tr\left( W_{\a} e^V \bar{W}^2 e^{-V} A e^V \bar{A} e^{-V}(AB)^k\right), \qquad
E_0 = \frac{3}{2}k + \frac{13}{2}, \quad r = k-1, \\
E^k &=& Tr\left( W^2 e^V \bar{W}^2 e^{-V} A e^V \bar{A} e^{-V} (AB)^k\right), \qquad
E_0 = \frac{3}{2}k + 8, \quad r = k, \\
F^k &=& Tr\left( e^V \bar{W}^2 e^{-V} A e^V \bar{A} e^{-V} (AB)^k\right), \qquad
E_0 = \frac{3}{2}k + 5, \quad r = k-2,\\
G^k &=& Tr\left(e^V \bar{W}_{\da} e^{-V} A e^V \bar{A} e^{-V} (AB)^k\right), \qquad
E_0 = \frac{3}{2}k + \frac{7}{2}, \quad r = k-1, \\
H^k &=& Tr\left( e^V \bar W_{\da} e^{-V}W^2 A e^V \bar{A} e^{-V} (AB)^k\right), \qquad
E_0 = \frac{3}{2}k + \frac{13}{2} , \quad r = k+1.
\eea
It must be noted that $G^k$ coincides with $O_{\da}^{1nk}$ for $n=1$ and
$H^k$ coincides with $O_\a^{2nk}$ for $n=1$.
Moreover, $D^k$ coincides with the operator $\bar O_\a^{2nk}$ for $n=1$ and $k=0$.

Inspection of the above list shows that these families are the lowest
dimensional operators of a given structure, with building blocks
 given by $W_{\a}$,
$A$, $\bar{A}$, $B$ and $\bar{B}$.

It should also be stressed that, although these operators have given quantum
numbers of $SU(2)\times SU(2)$, and of $SU(2,2|1)$ $E_0$, $s_1$, $s_2$, $r$,
we have not discussed
the most general form of these operators due to further mixing in terms of
the constituent singleton fields $W_\alpha$, $A$, $B$.
For instance, we have not written terms involving $D_\alpha A$ or $D_\alpha B$,
which certainly occur in the completion of some of the above operators
(For example the ones including $J_{\a\da}^k$  which contain both $W_{\a}\bar{W}_{\da}$
and $D_{\a}A\bar{D}_{\da}\bar{A}$ (or $A \leftrightarrow B$)).

Finally, we analyse the \eqn{CC} condition.
In this case the only multiplet which does not violate the \eqn{addi} 
inequality  is the type III 
vector multiplet, for which we get $E_{0} = \dfrac{3}{2}r + 2$.  
This apparently could be interpreted as shortening to a semilong 
vector multiplet.
However, the states of such multiplet do not appear in the KK 
expansion, while the states which are  
complementary to them form a chiral hypermultiplet which is allowed 
by the KK analisys\footnote{Physically, the exclusion of the semilong multiplet 
can also be seen by the fact that it would contain an additional massless vector 
for $j=l=r=0$ which do not correspond to any symmetry besides the 
isometry and barion symmetry.}.
Its lowest state is 
the $\phi$ field with $E_{0}^{(s)} = E_0 + 1 = \dfrac{3}{2} r^{(s)}$, 
which is indeed the group theoretical condition for the shortening to 
a chiral multiplet of the type given in \eqn{Sk}. 
The $k=0$ ($r^s = 2$) chiral multiplet has as last component a 
complex massless scalar related to the $A_{ab}$ 2--form wrapped on 
the non--trivial 2--cycle of $T^{11}$, giving a second complex 
modulus other than the dilaton $B$ for type IIB on $AdS_5 \times 
T^{11}$.
Note that there is another massless scalar in the serie $S^k$ \eqn{Sk}for 
$k=2$.
This corresponds to the spin $j=l=1$ in the harmonic expansion in the 
internal metric $h_{ab}$.

\medskip

We would also like to  remark that there are many more operators in the
gauge theory which do not correspond to any supergravity KK mode, even
though these multiplets may have spin less than two.
A typical example is the Konishi (massive vector) superfield \cite{Kon}
\beq
K= Tr(A e^V \bar{A} e^{-V}) + Tr(B e^V \bar{B} e^{-V})
\eeq
with $r=0$ and in the $G$--singlet $j=l=0$.

This superfield has anomalous dimension \cite{ans}.
However, inspection of the supergravity spectrum, shows that the 
multiplets with $j=l=r=0$
must have rational dimension and indeed they were identified with
$Q^{k=0}$ in \eqn{nonpro} with $E_0 =6$ and the Betti multiplet ${\cal U}$
in \eqn{cU} with $E_0 = 2$.

This state of affair is resolved by the fact that $K$ is expected to have
a divergent dimension $\Delta$ in the large $N$--limit, as presumably
happens in the $\cN = 4$ theory so that it should correspond to a
string state.

The Konishi multiplet \cite{Kon}
is a long multiplet whose $\bar{D}^2$ is a chiral
superfield which is a linear combination of the superpotential
${\cal W} = \e^{ij}\e^{kl}Tr(A_i B_k A_j B_l)$ and $Tr(W^{\a}W_{\a})$.
This implies that neither ${\cal W}$ nor $Tr(W^{\a}W_{\a})$ are chiral
primaries but rather a combination orthogonal to $\bar{D}\bar{D} K$.
It is the latter superfield which appears in the supergravity spectrum
and coincides with the chiral dilaton multiplet $\Phi^k$ with $k=0$.
This  is an example of operator mixing alluded before.

Finally we observe that the knowledge of the  flavour and $R$--symmetry
anomalies in the  gauge theory allow one to completely fix the low energy
effective action of Type IIB supergravity on $AdS_5 \times T^{11}$ at least
in the sector of the massless vector multiplets \cite{FZ}. In fact this relies
on the computation of the bulk Chern--Simons term of the several gauge
factors involved \cite{GST}
\beq
d_{\Lambda\Sigma\Delta} \int F^\Lambda \wedge F^\Sigma \wedge A^\Delta \ .
\eeq
where $\Lambda=1,\ldots , 8$ with $U_R(1)$, $U_b(1)$ and 
$SU_A(2)\times SU_B(2)$ gauge factors.

Because of the $AdS/CFT$ correspondence, the gauge variation of such 
Chern--Simons terms must precisely match, at least in leading order in $N$,
the current anomalies of the boundary gauge theory \cite{W,FZ,HeSk,G,AnKe}. 
Moreover the mixed gravitational gauge Chern--Simons terms
\beq
c_\Lambda \int A^\Lambda\wedge Tr\ R\wedge R ,
\eeq
(where $\Lambda$ here runs only over the $U(1)$ factors of the bulk gauge 
fields)
should be non--leading since they are related to string corrections in the
$AdS/CFT$ correspondence \cite{AnKe}. Because of the particular matter 
content of the model \cite{KW}, all coefficients are in principle
 proportional to $N^2$ and thus leading in the $AdS/CFT$ 
duality.

So it is crucial that $c_\Lambda=0$, {\sl i.e.} that $U_R(1)$, $U_b(1)$ are
traceless\cite{G}.
 The only non--vanishing $d_{\Lambda\Sigma\Delta}$ coefficients
are 
\beq
d_{rAA}=d_{rBB},\qquad d_{bAA}=-d_{bBB}, \qquad d_{rrr},\qquad d_{rbb}
\eeq
and thus they determine (up to two derivatives) the low energy effective
action.

\vskip 1truecm
\paragraph{Acknowledgements.}
\ We would like to thank I. Klebanov, M. Porrati and especially A. Zaffaroni for
discussions. A.C. is grateful to CERN for the kind
 hospitality during the early stages of this work. \\
This research is supported in part by EEC under TMR
contract ERBFMRX-CT96-0045 (Politecnico di Torino, LNF Frascati,
LPTENS Paris) and by DOE grant DE-FGO3-91ER40662 .

\section*{Appendix A: Notations and Conventions}
\setcounter{equation}{0}
\makeatletter
\@addtoreset{equation}{section}
\makeatother
\renewcommand{\theequation}{A.\arabic{equation}}
Consider $AdS_5 \times T^{11}$.
We call $M,N$ the curved ten--dimensional indices,
$\mu,\nu$/$m,n$
the curved/flat $AdS_5$ ones and  $\a\beta$/$a,b$ the curved/flat
$T^{11}$ ones.
In the four dimensional CFT $\a,\ldots$ and $\da,\ldots$
are spinorial indices.

Our ten--dimensional  metric is the mostly minus $\eta = \{+-\ldots-\}$,
so that the internal space has a negative definite metric.
For ease of construction, we have also used a negative metric to raise
and lower the $SU(2)\times SU(2)$ Lie--algebra indices.

Furthermore, for the $SU(2)$ algebras we have defined $\e^{123} = \e^{12} = 1$.

The $SO(5)$ gamma matrices are
\beq
\gamma_1 = \left(
\bet{cccc}
&&&1 \\
&& 1 & \\
& -1 && \\
-1 &&&
\eet
\right) \qquad
\gamma_2 = \left(
\bet{cccc}
&&&-i \\
&& i & \\
& i && \\
-i &&&
\eet \right)
\eeq
\beq
\gamma_3 = \left(
\bet{cccc}
&&1& \\
&&&-1 \\
-1&&& \\
 &1&&
\eet \right) \qquad \gamma_4 = \left(
\bet{cccc}
&&i& \\
&&&i \\
i& && \\
 &i&&
\eet \right)
\eeq
\beq
\gamma_5 = \left(
\bet{cccc}
i&&& \\
&i&& \\
&&-i& \\
&&&-i
\eet \right)
\eeq

%
%

\end{document}